\documentclass[acmtog]{acmart}
\acmSubmissionID{338}
\usepackage[ruled,vlined]{algorithm2e}
\RestyleAlgo{ruled}
\RestyleAlgo{vlined}

\renewcommand\footnotetextcopyrightpermission[1]{} % removes footnote with conference information in first column

\acmJournal{TOG}

\usepackage{booktabs}
\usepackage{subfigure}

\newcommand{\bomega}{{\mathbf{\omega}}}

\SetKwComment{Comment}{// }{}
\makeatletter
\newcommand\paragraphNew{\@startsection{paragraph}{4}{\parindent}%
  {-.5\baselineskip \@plus -2\p@ \@minus -.2\p@}%
  {-3.5\p@}%
  {\ACM@NRadjust{\@parfont}}}
	\makeatother
	
\AtBeginDocument{%
  \providecommand\BibTeX{{%
    \normalfont B\kern-0.5em{\scshape i\kern-0.25em b}\kern-0.8em\TeX}}}

\usepackage{multirow} % For merging cells

\acmJournal{TOG}

\begin{document}


\title{A Micro-Ellipsoid Model for Wet Porous Materials Rendering }
%\title{WetSpongeCake: a Surface Appearance Model Considering Porosity and Saturation}

% \title{Height-free Multiple-bounce Smith Microfacet BRDFs}
%multiple-bounce, path space.
% Shadowing-masking

\author{Gaole Pan}
\affiliation{%
  \institution{Nanjing University of Science and Technology}
  %\city{Nanjing}
  \country{China}
}
\email{pangaole@njust.edu.cn}

\author{Yuang Cui}
%\authornote{Joint first authors.}
\orcid{0009-0006-8983-7844}
\authornote{Research done when Yuang Cui was an intern at Nanjing University of Science and
Technology. }
\affiliation{%
  \institution{Anhui Science and Technology University}
  %\city{Bengbu}
  \country{China}
}
\email{yuangcui@outlook.com}

\author{Jian Yang}
\affiliation{%
  \institution{Nanjing University of Science and Technology}
  %\city{Nanjing}
  \country{China}
}
\email{csjyang@njust.edu.cn}

\author{Beibei Wang}
\authornote{Corresponding author.}
%\thanks{$^\dagger$Corresponding author.}
\affiliation{
    \institution{Nanjing University}
    \country{China}
}
\email{beibei.wang@nju.edu.cn}
%\affiliation{%
%  \institution{Nankai University}
%  \city{Tianjin}
%  \country{China}
%}
%\affiliation{%
%  \institution{Nanjing University of Science and Technology}
%  \city{Nanjing}
%  \country{China}
%}

\renewcommand{\shortauthors}{Pan et al.}

\begin{abstract}

Wet porous materials, like wet ground, moist walls, or wet cloth, are common in the real world. These materials consist of transmittable particles surrounded by liquid, where the individual particle is invisible in the macroscopic view. While modeling wet porous materials is critical for various applications, a physically based model for wet porous materials is still absent. In this paper, we model these appearances in the media domain by extending the anisotropic radiative transfer equation to model porosity and saturation. Then, we introduce a novel particle model- \emph{micro-ellipsoid}- by treating each particle as a transmittable ellipsoid, analogous to a micro-flake, to statistically characterize the overall optical behavior of the medium. This way, the foundational theory for media with porosity and saturation is established. Building upon this new medium, we further propose a practical bidirectional scattering distribution function (BSDF) model within the position-free framework--\emph{WetSpongeCake}. As a result, our WetSpongeCake model is able to represent various appearances of wet porous materials using physical parameters (e.g., porosity and saturation), allowing both reflection and transmission. We validated our model through several examples: a piece of wet cloth, sand saturated with different liquids, or damp sculptures, demonstrating its ability to match real-world appearances closely.

\end{abstract}


%\keywords{microflake, layered BSDF, multiple scattering}
% \keywords{microflake, layered BSDFs, position-free, multiple scattering}

\begin{teaserfigure}
\centering
\includegraphics[width=\textwidth]{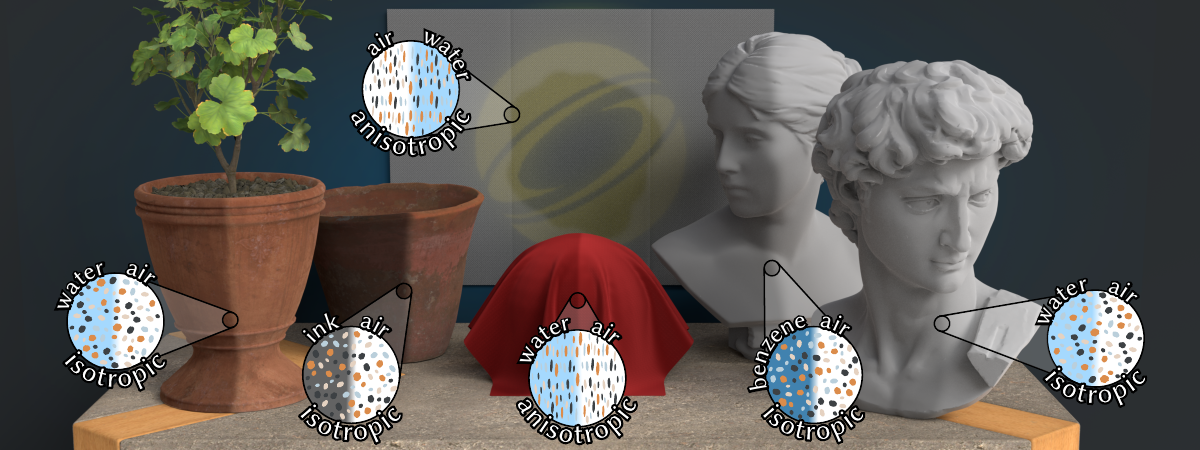}
\caption{ In this paper, we present a transmittable micro-ellipsoid model to render wet porous materials with physical parameters. It can produce vivid appearances on a wide range of materials. Examples include the wet cloths, flowerpots and sculptures as shown above. 
% \added{we want to talk about the anisotropic or isotropic particles in this figure. The reviewer might miss it.}
%\added{this is placeholder. similar to previous teaser, with a bunch of stuff on the table, and some of them are wet. this is going to be a nice teaser! use maps to specify the parameters.} }
}
\label{fig:teaser}
\end{teaserfigure}

\maketitle

\section{Introduction}
%sand clay powered materials.  filled with air. some interesting finds: becomes darker, and translucent. ~\cite{Jensen1999} the average scattering angle is reduces, leads to more absorption. 

Reproducing appearances from the real world with material models is essential in computer graphics. Material models for some common appearances (e.g., plastic, metals, glasses, etc.) have been extensively developed with surface shading models (e.g., microfacet models~\cite{Cook1982REFLEC, walter2007mmrt}). Unfortunately, these widely-used shading models have limited capability to represent \emph{wet porous materials}. Examples of this kind of materials include a damp road or wet sand (see Fig.~\ref{fig:sand}). Generally, the porous materials are made of many tiny grains. After being filled with liquids, the materials become wet and appear darker. These appearances have become crucial for many applications, like wet roads in the autonomous driving simulation. However, it's challenging to simulate these materials, as it involves complex optical phenomena. %Therefore, in this paper, we seek a practical surface shading model for wet-powdered materials. 

In the literature, wet porous materials can be represented at different levels--geometry, medium or surface materials. While modeling individual particle as a sphere~\cite {Peltoniemi1992} or an prolate spheroid~\cite{Kimmel:07} is accurate, it leads to extensive memory and time cost. Another group of methods render wet or porous materials with radiative transfer equation (RTE) or its variant. However, they model saturation or porosity of medium in an ad-hoc way, by tweaking the Henyey-Greenstein (HG) phase function parameters~\cite{Jensen1999} or unnormalized porosity-aware transmittance functions~\cite{Hapke2008}. Either way leads to unrealistic rendering. To our knowledge, no foundational medium formulation considering both porosity and saturation exist.
The final category of methods represents porous materials using a bidirectional reflectance distribution function (BRDF) \cite{Merillou2000, Hnat2006, Jianye2006}, incorporating cylinder-shaped holes on the surface. While these approaches are practical, they struggle to accurately reproduce real-world effects and cannot simulate transmission phenomena.
%For instance, a small difference in the refractive index of liquids can cause a significant variation in perceived darkness (see Fig.~\ref{fig:sand}), which cannot be captured by the aforementioned surface models.

In this paper, we aim to model a wet porous appearance with controllable physical parameters that closely replicate real-world effects while maintaining practicality. To this end, we propose a generalized anisotropic radiative transfer equation to model a wet porous medium considering both the porosity and saturation effect. At the core of our wet porous medium is a transmittable micro-ellipsoid model, analogous to the micro-flake model~\cite{jakob2010radiative}, except with different particle shapes. 
In our model, each particle is represented with an ellipsoid, and the distribution of aggregated particles follows an ellipsoidal distribution. We derive phase function and anisotropic attenuation functions for both isolated and aggregated medium particles. While the generalized anisotropic RTE, along with the transmittable micro-ellipsoid model, allows rendering a wet porous medium, it requires a long convergence time. To address this issue, we further propose a practical surface appearance model within the position-free framework~\cite{Guo:2018:Layered,Wang:2022:sponge}--\emph{WetSpongeCake}. 
%Our WetSpongeCake model includes both single scattering and multiple scattering of micro-ellipsoids, where the former is \added{computed analytically} and the latter is mapped a modified single scattering parameters using a small neural network.
Consequentially, our WetSpongeCake model can faithfully reproduce wet porous materials with physical parameters (porosity and saturation). It can represent a wide range of appearances, including a piece of wet paper, saturated sand, or cloth. To summarize, our contributions include:
\begin{itemize}
    \item a generalized anisotropic radiative transfer equation for wet porous medium, which can capture the effect of surrounding liquid on light propagation through particles,
    \item a micro-ellipsoid model for wet medium particles, which defines its phase function and attenuation functions, and
    \item  a practical surface appearance model \emph{WetSpongeCake} to represent wet porous materials with physical parameters.
\end{itemize}

\section{Related Work}
\label{sec:related}
%-----------------------------------%

\subsection{Wet material models} 
%\added{discrete RTE}

Several groups of approaches have been proposed in the literature for representing wet materials, explicit simulations on randomly generated particles, medium-based models and surface-based BRDFs.

Methods in the first category represent individual particle as a sphere~\cite{Peltoniemi1992} or a prolate spheroid~\cite{Kimmel:07}, and then perform Monte Carlo simulations on randomly generated particles. These methods are computationally expensive and designed for specific material (e.g., sand). 
% Similar to Kimmel et al.~\shortcite{Kimmel:07}, we model each particle as a spheroid. However, instead of explicit simulating random walks among the particles, we simplify the light transport within each particle using a micro-phase function and derive a lightweight BSDF.

In the second category, Jensen et al.~\shortcite{Jensen1999} capture the saturation of materials using a combined surface and subsurface model, leveraging a two-term Henyey-Greenstein (HG) phase function to approximate the scattering of particles. While their method produces convincing results, it lacks physically meaningful parameters, requiring manual adjustments of the HG parameter for control. Meanwhile, other studies~\cite{Hapke2008, HAPKE1999565, SHKURATOV1999235} introduce a porosity parameter by modifying the original RTE. However, these medium-based approaches rely on volumetric path tracing, which results in significant computational costs. Additionally, several other methods~\cite{Moon2007, Meng2015, muller2016} have been proposed for simulating light transport in visible discrete particles, but these fall outside the scope of our work.

In the third category, several studies represent wet materials using surface-based models, including Lekner and Dorf~\shortcite{Lekner:88} and Bajo et al.~\shortcite{bajo2021physically}. The former is based on Angstrom's model~\shortcite{Anders1925}, while the latter improves upon Lekner et al.'s model~\shortcite{Lekner:88}. Other real-time surface models~\cite{Hnat2006, Merillou2000, Jianye2006} introduce a cylindrical shape for holes and apply intrinsic roughness to the hole surfaces. These approaches are lightweight and practical. However, they struggle to closely match real-world effects. As pointed out by Twomey et al.~\shortcite{Twomey:86}, even a minor refractive index (IOR) difference (e.g., water vs. benzene on wet sand~\cite{Craig1983}) can cause significantly different appearances, which surface models cannot accurately capture. More recently, d'Eon proposed several shading models~\cite{d2021analytic, d_Eon_VMF} to describe porous, diffuse-like BRDFs. While these models provide analytic solutions for porous materials, they do not account for the saturation effect. %\added{
Lucas et al. \shortcite{Lucas2023, lucas2024fully} model porous layers, such as dust or dirt, where grains are distributed on a surface. In contrast, our work models porous materials, such as sand or walls, where particles are distributed throughout the volume.
%While both approaches address porous structures, they tackle fundamentally different configurations.}
% \added{we need a deep discussion: what's the difference and similarity? }

%However, these methods have difficulty matching the real-world effects closely and are limited to reflection only. For example, an interesting observation is that different liquids with direct refraction index lead to different darkness that the surface model can not capture.

%Therefore, we need to consider the media.
%Jensen et al.~\shortcite{Jensen1999} propose a specular for reflection and subsurface for the media, using two HG functions to approximate the wet appearance without any porosity parameters. They tweak the value of g to make the scattering more forward. Hapke~\shortcite{Hapke2008} improved the RTE, considering the porosity. When the porosity is 1, the medium comes back to the regular medium.

\subsection{Position-free BSDFs}
The position-free property was first introduced in volumetric BSDF models by Dupuy et al.~\shortcite{Dupuy:2016:Unification}, which start the trend of volumetric BSDF models. Guo et al.~\shortcite{Guo:2018:Layered} reformulate path integral within a medium from the entire spatial dimensions to the depth dimension only, as the incident and exit rays share the same position. This simplification of the path integral leads to a more efficient BSDF evaluation. This idea has been used further by Xia et al.~\shortcite{Xia:2020:Layered} and Gamboa et al.~\shortcite{Gamboa:2020:EfficientLayered} for more advanced sampling approaches or modeling pearlescent pigments~\cite{Guillen2020Pearlescent}. 
Wang et al.~\shortcite{Wang:2022:sponge} further generalize single-scattering to handle finite-thickness slabs with transmission. Their SpongeCake model simplified the computation by treating materials as volume slabs without surface interfaces. They also approximated multiple scattering using single scattering with adjusted parameters, making BSDF evaluations efficient and noise-free. Our surface appearance model also follows the position-free framework as a practical solution for wet porous medium rendering.

% In our method, we adopt this position-free framework by integrating the medium into the SpongeCake model. For multiple scattering, we utilize the parameter mapping technique from SpongeCake model to maintain efficiency and accuracy.

\subsection{Micro-flake model}
The micro-flake model, introduced by Jakob et al.~\shortcite{jakob2010radiative}, was created to capture the complex anisotropic behavior of participating media by modeling the distribution of reflective micro-flakes. Later, Heitz et al.~\shortcite{heitz2015SGGX} introduced the symmetric GGX (SGGX) representation, which uses $3\times3$ positive-definite matrices to efficiently describe micro-flake distributions. This method is flexible, allowing for both surface-like and fiber-like micro-flakes, with easy control over their orientation. However, the micro-flake model treats each particle as an ideal disk, which limits its ability to represent transmittable particles.

\begin{table}[!t]
	\renewcommand{\arraystretch}{1.1}
	\caption{\label{tab:notations} Notations.}
\begin{small}
 % \left
  \begin{tabular}{|l|c|l|}\hline
  \multicolumn{2}{|c|}{Mathematical notation }\\
	\hline
$\Omega$ &full spherical domain \\  
$\cos\bomega$ & The cosine of the angle between $\bomega$ and the z-axis \\
$\bomega_i \cdot \bomega_o$ &dot product \\
\hline
\multicolumn{2}{|c|}{Physical quantities}\\
\hline
$\omega_i $ & incident direction \\
$\omega_o $ & outgoing direction \\
$f_p(\theta)$& phase function\\
\hline
\multicolumn{2}{|c|}{Phase function parameters}\\
\hline
$\eta_p$ & refractive index of particle\\
$\sigma_p$& particle projected area\\
% $\Psi$ & sphericity\\
% $R$ & roughness\\
\hline
\multicolumn{2}{|c|}{Medium parameters}\\
\hline
$\sigma$ &normal projected area\\
$Z$ & thickness\\
$S$ & saturation\\
$P$ & porosity\\
% $n$ & particle count per unit volume\\
$\alpha$ & single-scattering albedo\\
% $\sigma_t$ & extinction coefficient\\
$\eta_l$ & refractive index of liquid\\
$\sigma_{t}^l$ & liquid extinction coefficient\\
\hline
\end{tabular}

\end{small}
\end{table}
%--------------------------%
\section{Wet porous materials with volume rendering}
%--------------------------%

\begin{figure}
\centering
\includegraphics[width = 0.8\linewidth]{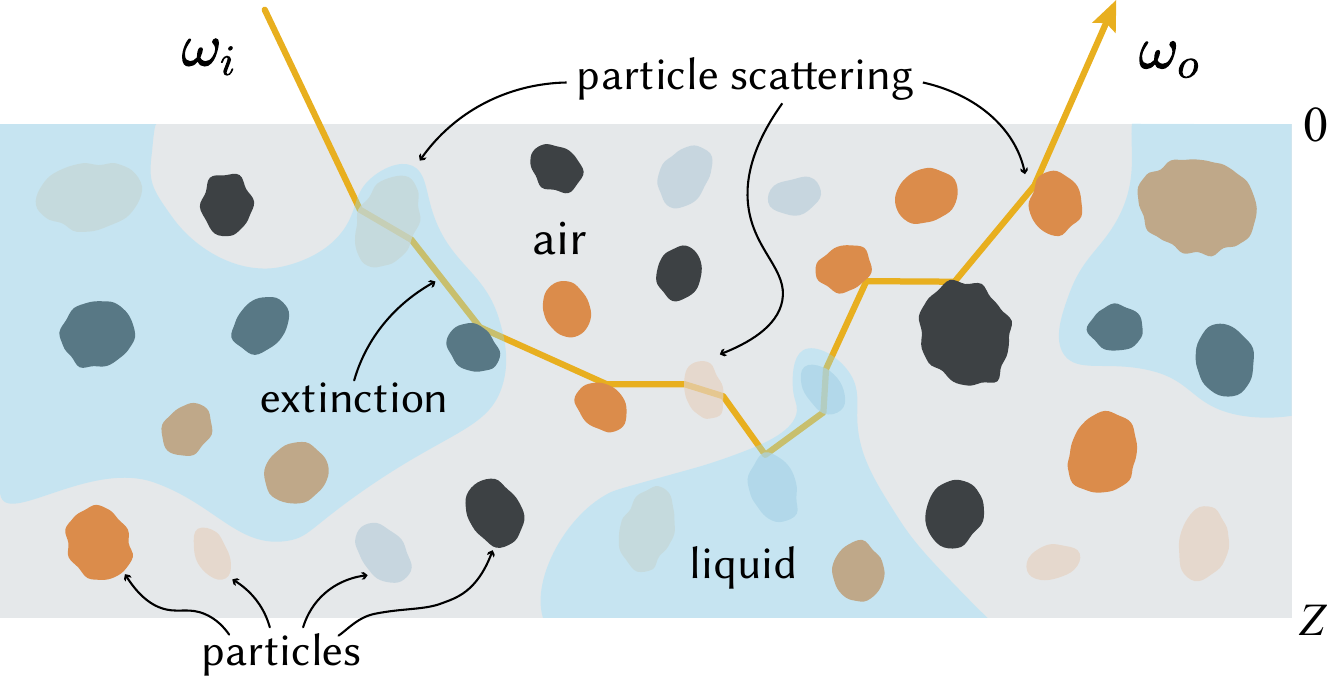}
\caption{Wet porous material is made of discrete particles, air, and liquid. After a light ray enters this material, it can be either reflected or refracted by the particle surface and travels in the air or the liquid. The ray is bounced within this material until it leaves the surface. }
\label{fig:config}
\end{figure}

In this section, we first introduce the underlying optical effects of wet porous materials (Sec.~\ref{sec:defination}). Next, we modify the radiative transfer equation to account for porosity and saturation (Sec.~\ref{sec:rte}). 
% Finally, we present our particle model, designed to capture the influence of liquid on scattering direction (Sec.~\ref{sec:micro-ellipsoid}).

\subsection{Wet porous materials}
\label{sec:defination}
%\added{Preliminaries moved to here}

Wet porous materials are composed of millions of non-spherical particles distributed in air and liquid, forming a complex volume, as illustrated in Fig.~\ref{fig:config}. The particles allow both reflection and transmission. These particles are characterized by intrinsic properties such as refractive index $\eta_p$ and albedo $\alpha$, while their orientations follow a specific distribution. The porosity, $P$, represents the fraction of the volume not occupied by particles. The liquid within the material has a refractive index $\eta_l$ and an extinction coefficient $\sigma_t^{l}$. The fraction of the liquid occupying the non-particle volume is defined as the saturation $S$. All parameters are summarized in Table~\ref{tab:notations}.

When a light ray enters the medium, it interacts with particle surfaces, where it is either reflected or refracted. Between interactions, the ray travels through air or liquid, with the liquid absorbing some of its energy. The ray continues to bounce within the material until it exits the volume. Due to the finite thickness ($Z$) of the volume, the ray can either be reflected (exiting on the same side as its entry) or transmitted (exiting on the opposite side). The refraction between air and liquid occurring within the medium is ignored, following previous work~\cite{Kimmel:07}. We do not consider diffraction effects, as the diffraction patterns in a densely packed particulate media are altered by the proximity of neighboring particles, making diffraction effects negligible, as discussed by Hapke~\shortcite{Hapke2008}.
% \added{Gaole: please make it more accurate. in the medium, right?}

% \added{about the diffraction, why we do not need to consider it? please check the Hapke paper.}
%\added{Diffraction effects are not considered in our model because, in densely packed particulate media, the diffraction patterns are altered by the proximity of neighboring particles. This makes traditional Fraunhofer diffraction patterns indistinguishable from the collimated incident light, making their contribution negligible in the reflectance model.}

\begin{figure}
\centering
\includegraphics[width = 0.75\linewidth]{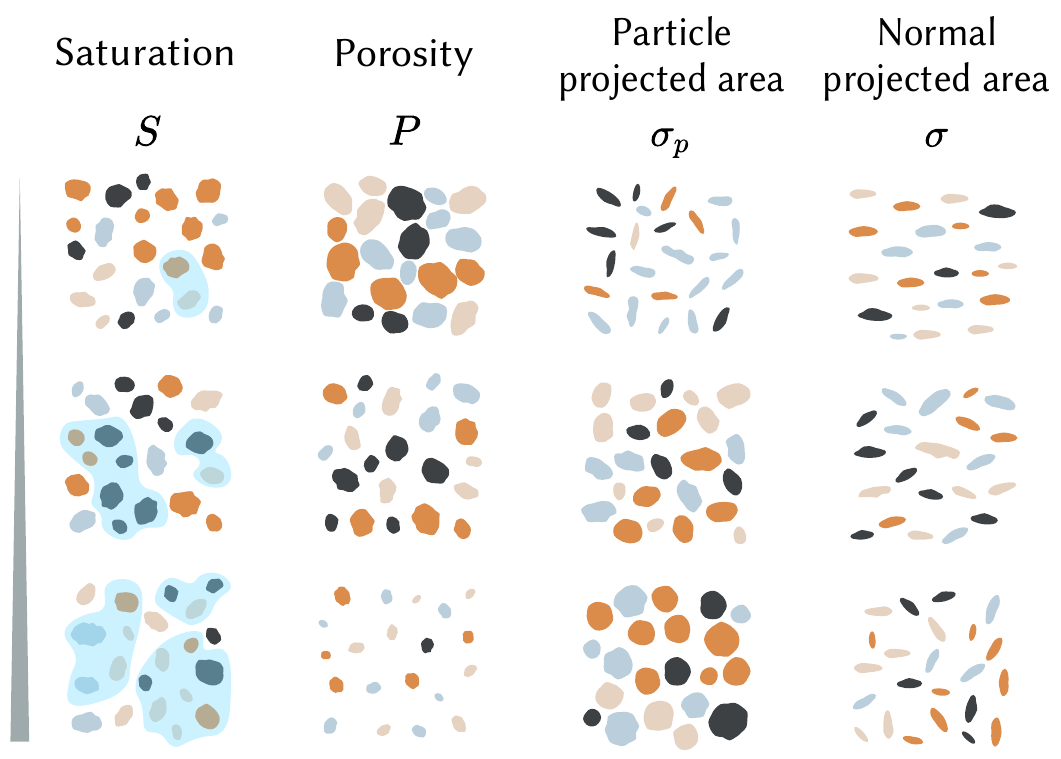}
\caption{Several examples of parameters defined on particles and medium.}
\label{fig:particleConfig}
\end{figure}

%--------------------------%
\subsection{Generalized RTE considering porosity and saturation}
\label{sec:rte}
%--------------------------%
The anisotropic RTE~\cite{jakob2010radiative} is formulated as:
\begin{equation}
    \begin{aligned}
(\omega \cdot \nabla) L(\omega)=&-\sigma_{t}(\omega) L(\omega)+\sigma_{s}(\omega) \int_{\Omega} f_{p}\left(\omega^{\prime} \cdot \omega\right) L\left(\omega^{\prime}\right) \mathrm{d} \omega^{\prime} \\
&+ Q(\omega), 
\label{eq:rte}
\end{aligned}
\end{equation}
where $\sigma_t$ and $\sigma_s$ are the extinction and scattering coefficients respectively. $Q$ is the source term. Note that the dependence on the position in Eqn.~(\ref{eq:rte}) is ignored for compactness. %\added{The self-emissive term is also omitted, as it's unnecessary in our case.} 

% Then, the transmittance for traveling a distance $t$ in anisotropic medium is given as: 
% \begin{equation}
% \label{eq:traditionalT}
% T(t,\omega) = e^{-\sigma_t(\omega) t}. 
% \end{equation}

The anisotropic radiative transfer equation allows for anisotropic scattering in the sense that the scattered intensity depends on the scattering angle. However, it does not take into account porosity and saturation, which limits its ability to represent wet porous materials. To enhance its representation capabilities, we incorporate both porosity and saturation into the anisotropic RTE.

\paragraph{\textbf{Porosity}}
Porosity describes the fraction of the medium that is not occupied by particles. Hapke~\shortcite{Hapke2008} incorporated porosity into the RTE by treating the discrete medium as a series of discrete layers of particles, resulting in a discontinuous stair-step transmittance. Then, the discontinuous transmittance is transformed into an equivalent continuous function as an approximation:
\begin{equation}
\label{eq:HapkeT}
    T(t) = K e^{-K \sigma_{t} t },
\end{equation}
where $K$ is a key factor related to the porosity, derived by considering the that adjusts both the initial value and the extinction rate of the transmittance. When the particles are large compared to the wave-length, $K$ is only affected by the porosity $P$, hence:
\begin{equation}
\label{eq:K}
K=-\frac{\ln{\left(1-\left(\frac{3\sqrt{\pi}}{4}\left(1-P\right)\right)^{\frac{2}{3} }\right)}}{\left(\frac{3\sqrt{\pi}}{4}\left(1-P\right)\right)^{\frac{2}{3} }},
\end{equation}
the detailed derivation is shown in the supplementary.

%\added{A Taylor series expansion} of $K$ reveals that $K>1$.% indicating that transmittance larger than 1. must start at 1 to ensure conservation of energy. indicating Hapke's model is not , 

While Eqn.~(\ref{eq:K}) models porosity, it can not ensure energy conservation, as the transmittance can exceed 1. Inspired by Hapke's work, we modify the anisotropic RTE~\cite{jakob2010radiative} to incorporate porosity in a simple yet energy-conserving manner by introducing the porosity-related factor $K$ into the coefficients:
\begin{equation}
\begin{aligned}
(\omega \cdot \nabla) L(\omega)=&-K\sigma_{t}(\omega) L(\omega)+K\sigma_{s}(\omega) \int_{\Omega} f_{p}\left(\omega^{\prime} \cdot \omega\right) L\left(\omega^{\prime}\right) \mathrm{d} \omega^{\prime}\\
&+Q(\omega),
\label{eq:porosity}
\end{aligned}
\end{equation}
where $K$ behaves like the density; as porosity decreases, $K$ increases, leading to a denser medium. In our formulation, the transmittance function starts at 1, ensuring energy conservation. 

\begin{figure}[t]
\centering
\includegraphics[width = 1.0\linewidth]{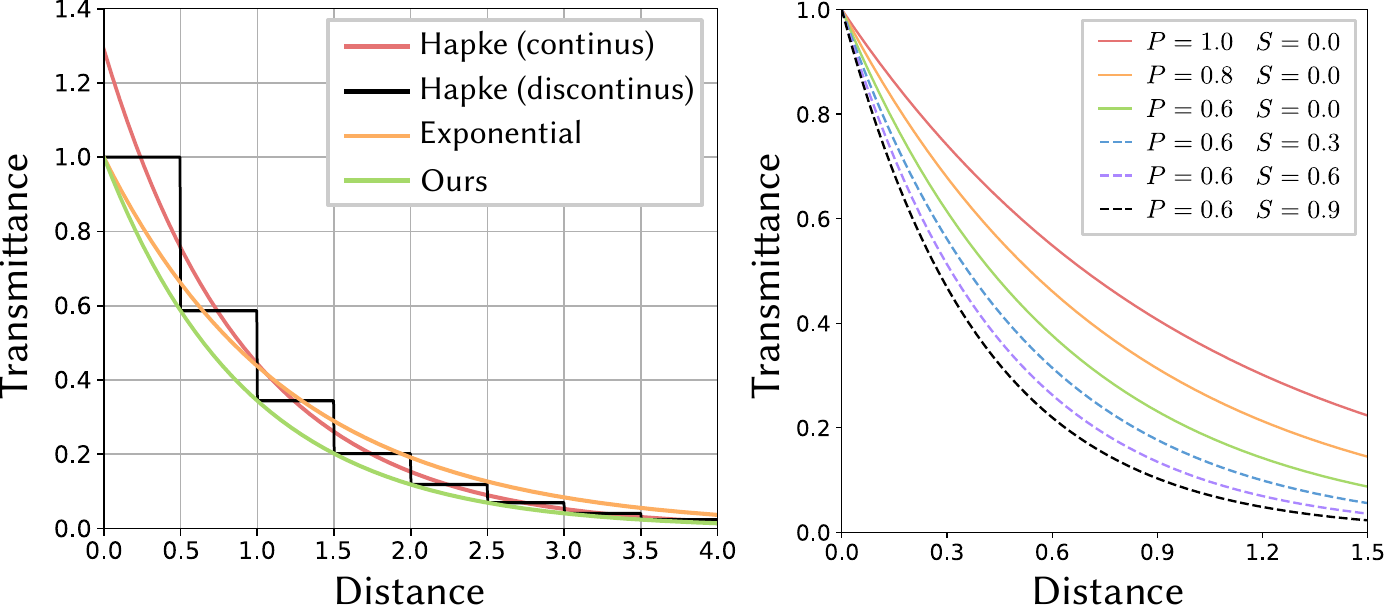}
\caption{Left: Comparison of several transmittance functions w.r.t. the traveling distance. Our saturation is set as 0. Right: our transmittance function as a function of the distance across several parameters.
% \added{dashed for S.}
}
\label{fig:transmittance}
\end{figure}

We compare Hapke's discontinuous stair-step transmittance, its continuous approximation, traditional exponential transmittance, and our transmittance in Fig.~\ref{fig:transmittance}. Clearly, Hapke's continuous approximation starts with an initial value greater than 1, which violates the principle of energy conservation. In contrast, our transmittance, $T(t) = e^{-\left(K \sigma_{t}\right) t} $, not only ensures energy conservation but also also takes porosity effects into account.

% \added{do we want to include the regular exponential function?}

%To ensure energy conservation, we do not model porosity's impact on the initial value of transmittance, focusing instead on its influence on the extinction rate.

% \begin{equation}
%     \begin{aligned}
% (\omega \cdot \nabla) L(\omega)=&-K\sigma_{t}(\omega) L(\omega)+K\sigma_{s}(\omega) \int_{\Omega} f_{p}\left(\omega^{\prime} \cdot \omega\right) L\left(\omega^{\prime}\right) \mathrm{d} \omega^{\prime},\\
% K=&-\frac{\ln{(1-1.209(1-P)^{2/3})}}{1.209(1-P)^{2/3}},
% \end{aligned}
% \end{equation}

% where $P$ is the porosity, and $K$ is computed as a function of porosity. Notably, $K$ behaves similarly to density; as porosity decreases, $K$ increases, resulting in a denser medium. \added{This relationship aligns with intuitive expectations.}

\paragraph{\textbf{Saturation}}
After porous materials are wetted by a liquid, the fraction of the liquid occupying the non-particle volume is defined as saturation $S$. This saturation also influences the anisotropic RTE. We assume that the liquid is an absorption-only medium. When the liquid permeates the medium, the distribution of particles remains unchanged, as the liquid simply replaces the air fraction. 
%Furthermore, interactions between air and liquid within the medium are not considered. 
Under these assumptions, light extinction in the medium is influenced by both the particles and the liquid. While the attenuation caused by the particles has been modeled in Eqn. (\ref{eq:porosity}), the additional attenuation resulting from the liquid must be considered as well. Therefore, the anisotropic RTE considering porosity and saturation can be formulated:
\begin{equation}
    \begin{aligned}
(\omega \cdot \nabla) L(\omega)=
&-(K\sigma_{t}(\omega)+S\sigma_t^l) L(\omega)\\
&+K\sigma_{s}(\omega) \int_{\Omega} f_{p}\left(\omega^{\prime} \cdot \omega\right) L\left(\omega^{\prime}\right) \mathrm{d} \omega^{\prime}+Q(\omega),
\end{aligned}
\end{equation}
where $\sigma_{t}^l$ is the liquid's extinction coefficient. The transmittance is obtained by integrating extinction of light along $\omega$, yielding:
\begin{equation}
\label{eq:ourT}
    T(t) = e^{-\left(K \sigma_{t} + S \sigma_{t}^l\right) t }.
\end{equation}

Note that when $P=1$ and $S=0$, the above transmittance reduces to the typical exponential transmittance. In Fig.~\ref{fig:transmittance} (right), we illustrate the transmittance as a function of distance for various materials with different porosity and saturation.

\begin{figure}[t]
\centering
\subfigure[ellipsoidal particle] {
\centering
\includegraphics[width = 0.46\linewidth]{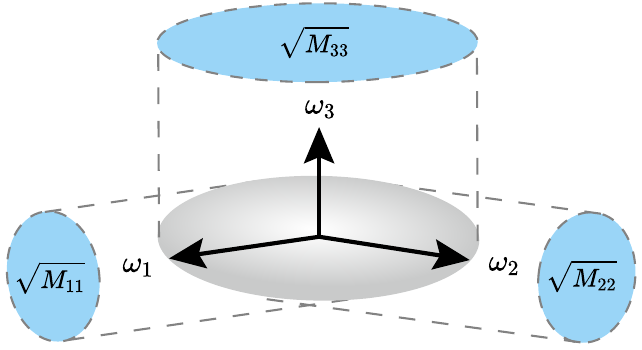}}
\quad
\subfigure[particle phase function] {
\centering
\includegraphics[width = 0.46\linewidth]{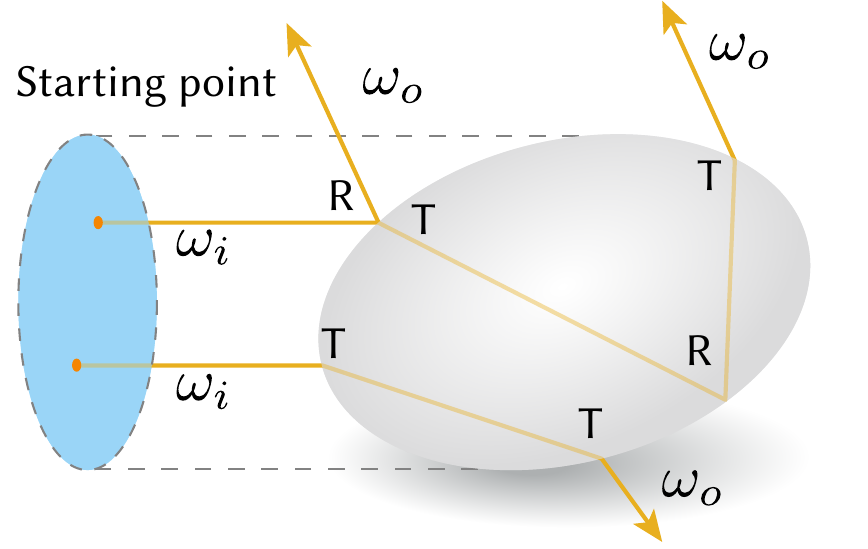}}
\caption{
(a) Each particle is modeled as an ellipsoid, defined by eigenvectors ($\omega_1$, $\omega_2$ and $\omega_3$) and eigenvalues, where $\omega_3$ is aligned with the ellipsoid normal. (b) the particle phase function is defined as the possibility of all paths originating from the projected area of $\omega_i$, traveling in direction $\omega_i$, bouncing within the ellipsoid, and finally exiting in direction $\omega_o$. }
\label{fig:ellipsoid}
\end{figure}

%——————————————————————————————————————————————————————%
\section{Transmittable Micro-ellipsoid model}
\label{sec:micro-ellipsoid}
%——————————————————————————————————————————————————————%

To render a wet porous medium with the generalized anisotropic RTE (Sec.~\ref{sec:rte}), we must establish the characteristics of the particles within the medium. Existing isotropic or anisotropic medium models have focused mainly on particles (micro-flake) that can reflect light and have not addressed transmittable particles. However, transmittability is essential for accurately representing a wet porous medium. In our model, we represent each particle as an ellipsoid, which allows for the modeling of an anisotropic medium while also enabling the transmittability of light. We will first define several key functions for isolated particles (Sec.~\ref{sec:individual}) and then present the statistical functions for aggregated particles (Sec.~\ref{sec:aggregated}).

%\remind{We aim for this particle model to have an analytical projected area, support a wide range of particle shapes, and account for the correlation between scattering direction and saturation. Ideally, it should also provide an analytical phase function. However, since achieving an analytical phase function is difficult for particles with volumes exhibiting non-delta micro-phase functions, we primarily focus on the first three goals and leave the last one for future work.}

%------------------------------------------------%
\subsection{Isolated ellipsoidal particle}
\label{sec:individual}
%------------------------------------------------%

Similar to the non-spherical particle by Jakob et al.~\shortcite{jakob2010radiative}, we need to define several functions to characterize a particle: 1) {projected area} $\sigma(\omega)$ is the area of the particle's projection onto $\omega^\perp$; 2) particle phase function $p(\omega_i,\omega_o)$ represents the probability of the light scattering into $\omega_o$, illuminated from direction $\omega_i$.

\paragraph{\textbf{Projected area}}

Each particle is represented as an ellipsoid with normal $m$, as shown in Fig.~\ref{fig:ellipsoid}. By setting orthonormal eigenvectors $(\omega_1,\omega_2,\omega_3)$, where $\omega_3 = m$, together with the projected area $\sigma_p$ onto $\omega_1$ and $\omega_2$, an ellipsoid can be established. It can be represented as a $3\times3$ symmetric positive definite matrix $M_p$:

%two We build a local coordinate for an ellipsoid by setting one axis \added{as $\omega_2 = m$}, and the others as $\omega_1$ and $\omega_3$, and they are orthonormal. 
%The projected area on to $\omega_1$ and $\omega_2$ are given by $\sigma_p$. This way, given an ellipsoid orientation, along with the projected area, an ellipsoid can be established. 

%it can be characterized by its projected areas and normal using a $3\times3$ symmetric positive definite matrix $M_p$:
\begin{equation}
\label{eq:matrix}
M_p = (\omega_1, \omega_2, \omega_3) \begin{pmatrix}
M_{11} & 0 & 0 \\
0 & M_{22} & 0 \\
0 & 0 & M_{33}
\end{pmatrix} (\omega_1, \omega_2, \omega_3)^T,
\end{equation}
where $M_{11}=\sigma^2(\omega_1)=\sigma_p^2$, $M_{22}=\sigma^2(\omega_2)=\sigma_p^2$ and $M_{33}=\sigma^2(\omega_3)=1$ are positive eigenvalues of $M$. %These eigenvalues correspond to the squared projected areas of the ellipsoid in the directions given by the orthonormal eigenvectors $(\omega_1,\omega_2,\omega_3)$. Specifically, $\omega_3$ represents the normal direction of the particle. \added{We model particle shape as a surface-like ellipsoid where eigenvalues are $(\sigma_p^2,\sigma_p^2,1)$.} 
The projected area of an ellipsoid with normal $m$ onto direction $\omega$ can be computed as:

\begin{equation}
\label{eq:projectedArea_SGGX}
\sigma(m, \omega) = \sqrt{\omega^T M_p \omega}.
\end{equation}

\paragraph{\textbf{Particle phase function}}
In contrast to the micro-flake model, which defines the particle phase function on a flat disk with a constant flake normal, our particle phase function is defined on an ellipsoid. Here, our particle phase function $p(\omega_i,\omega_o)$ represents the probability of all light paths exiting in direction $\omega_o$, given an entry direction $\omega_i$ with all possible entry points located on the ellipsoid surface. For a given entry point $x_i$, the light might have various types of light paths, denoted by $R$, $TT$, $TRT$, \dots, where $R$ stands for reflection and $T$ stands for transmission, resulting in a probability: 
% \revise{try to remove the dots}
\begin{align}
    p_i(x_i) = \sum_{t \in \mathcal{T}} A_t
	% p_i(x_i)=A_{R}+A_{TT}+A_{TRT}+A_{TRRT}+\dots.
 \label{eq:specular}
\end{align}
Here, \( A_t \) indicates the attenuation stemming from the Fresnel term along a specular path of type \( t \in \mathcal{T} \), where \( \mathcal{T} \) represents the set of all possible combinations of \( R \) and \( T \).
% Here, $A_t$ indicates the attenuation stemming from the Fresnel term along a specular path of type $t\in\{R,$ $TT,$ $TRT,$ $TRRT,$ $\dots\}$. 
%defined as:
%\begin{align}
%    &A_{t}=\delta(\mathbf{x}_n, \mathbf{x}_o, \omega_o)\times\notag\\
%    &(1-F(\mathbf{x}_0))(1-F(\mathbf{x}_n))(\Pi_{i=1}^{n-1}F(\mathbf{x}_i)),
% \label{eq:specularpaths}
%\end{align}

Then, we need to consider all the possible entry points $x_i$ on the ellipsoid surface. To do this, we equivalently transform it into all light paths starting from the projected area of $\omega_i$. This way, the particle phase function can be formulated as
\begin{equation}
    p(\omega_i,\omega_o) = \int_{\sigma(\omega_i)} p_i(x_i)\text{d}A x_i,
    \label{eq:microphase}
\end{equation}
where $\text{d}A x_i$ is the differential area around $x_i$.

Deriving a closed-form expression for the particle phase function is quite challenging. To address this, we estimate Eqn.~(\ref{eq:microphase}) with Monte Carlo simulation by sampling the projected area of $\omega_i$, shooting a ray at that direction, intersecting the ray with the ellipsoid to determine an entry point $x_i$. From $x_i$, the following specular paths are also sampled by choosing reflection or refraction at each interaction. When the ray exits the ellipsoid, we record its exit direction. This way, we tabulate the particle phase function as a 3D lookup table. Note that the one-dimension reduction is due to the azimuthal symmetry of the particle shape about the normal. For more detailed information, please refer to supplementary material.
% \remind{Sec.~\ref{sec:implementation}.}

%the closer the IOR of the particle and liquid are, the more likely the ray is to refract at intersections, resulting in a more forward-scattering micro-phase function. Hence, the micro-phase function encodes the influence of the surrounding liquid on the light’s scattering direction.

%represents the phase function of an individual particle used to determine the reflected direction $\omega_o$ of the ray and is formulated as:
%\begin{equation}
%\label{eq:microphase}
%    p(\omega_i,\omega_o) = \int_{\mathcal{P}} f\left( \Bar{x} \right) \, d\mu(\Bar{x}),
%\end{equation}
%where $\mathcal{P}$ is the space of paths, $\Bar{x} = (\omega_i,\ldots,\omega_o)$ is a path within the path space and $f\left( \Bar{x} \right)$ is the contribution of path $\Bar{x}$. 

\begin{figure}[t]
\centering
\includegraphics[width = \linewidth]{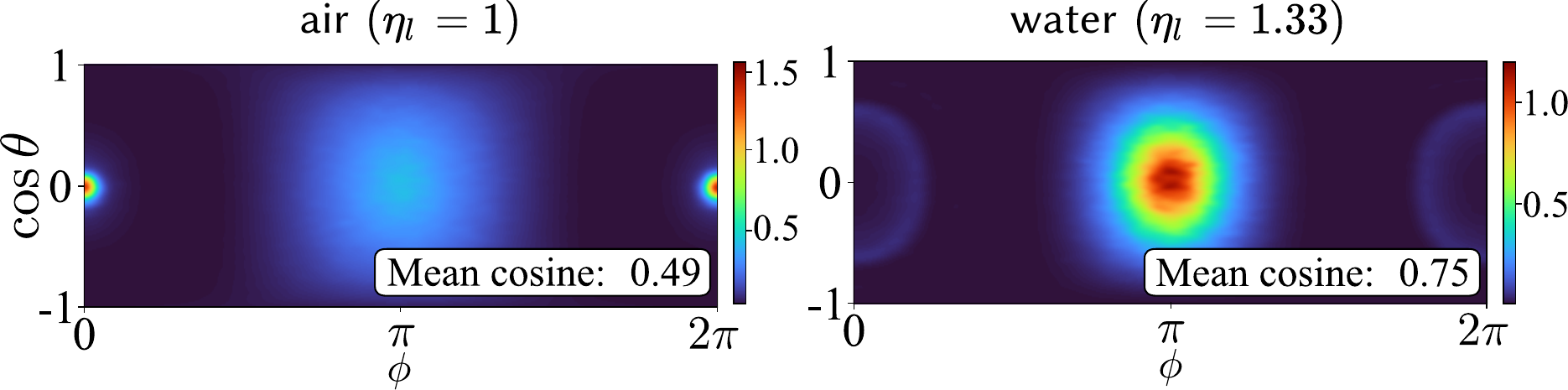}
\caption{Visualization of the phase function for particles surrounded by air and water, respectively, with the mean cosine value provided in the bottom-right corner. The remaining parameters are set to $\eta_p = 1.8$, $\sigma = 1$, and $\sigma_p = 1$. Additional results are provided in the supplementary material.}
\label{fig:phaseVis}
\end{figure}

% \begin{figure}[t]
% \centering
% \includegraphics[width = 0.6\linewidth]{./fig/flakeVsVolume.pdf}
% \caption{Comparison between the micro-ellipsoid model and the micro-flake model. Despite sharing the same normal distribution, the two media exhibit different projected areas, \added{primarily due to the differing projected areas of the respective particle types.}}
% \label{fig:flakeVsVolume}
% \end{figure}

%------------------------------------------------%
\subsection{Micro-ellipsoid model}
\label{sec:aggregated}
%------------------------------------------------%

Built on the isolated particle, we propose our micro-ellipsoid model, which is statistically defined for aggregated ellipsoidal particles. We assume the normal distribution of these aggregated ellipsoidal particles follows an ellipsoidal distribution, similar to the SGGX model~\cite{heitz2015SGGX}:
\begin{equation}
\label{eq:D_SGGX}
D(\omega_m) = \frac{1}{\pi \sqrt{|M_n|} (\omega_m^T M_n^{-1} \omega_m)^2},
\end{equation}
where $M_n$ is the symmetric positive definite matrix to describe the normal distribution, and $\omega_m$ is the micro-ellipsoid normal.  

\paragraph{\textbf{Attenuation coefficients}}
Next, we present another key function: the attenuation coefficients $\sigma_t(\omega)$ and $\sigma_s(\omega)$. Jakob et al.~\shortcite{jakob2010radiative} provided a general formulation for these functions:
%The volume extinction coefficient $\sigma_t(\omega)$ and scattering coefficient $\sigma_s(\omega)$ of a micro-ellipsoid medium are given by:
\begin{equation}
\label{eq:micro_volume_sigma}
\begin{aligned}
    \sigma_t(\omega) &= \int_{\Omega} \sigma(\omega_m, \omega) D(\omega_m) \, d\omega_m ,\\
    \sigma_s(\omega) &= \alpha \sigma_t(\omega).
  %  f_p(\omega_i , \omega_o) &= \frac{1}{\sigma_t(\omega_i)} \int_{\Omega} p(\omega_m, \omega_i , \omega_o) \sigma(\omega_m, \omega_i) D(\omega_m) \, d\omega_m.
\end{aligned}
\end{equation}
By applying the particle projected area function Eqn.~(\ref{eq:projectedArea_SGGX}), we have formulation of our attenuation coefficients.

\paragraph{\textbf{Phase function}}
It is essential to consider the medium's saturation, as it influences the IOR, which is a key factor in the phase function. With liquid in the medium, there are two cases of particle interface: air-particle and liquid-particle. We define the phase function for these two cases ($p_d$ and $p_w$), and then interpolate them w.r.t. the saturation:
\begin{equation} 
\begin{aligned}
    f_p(\omega_i,\omega_o) = 
    &\frac{1-S}{\sigma_t(\omega_i)} \int_{\Omega} p_d(\omega_m, \omega_i , \omega_o) \sigma(\omega_m, \omega_i) D(\omega_m) \, d\omega_m +\\
    &\frac{S}{\sigma_t(\omega_i)} \int_{\Omega} p_w(\omega_m, \omega_i , \omega_o) \sigma(\omega_m, \omega_i) D(\omega_m) \, d\omega_m.
\end{aligned}
\end{equation}

We now have the formulations for both the attenuation coefficients and the phase function, although neither has analytical solutions. Therefore, in practice, we compute these functions using Monte Carlo estimation and store the results in a 3D lookup table for the phase function and a 1D lookup table for the attenuation coefficients, as explained in supplementary material.

% \remind{Sec. \ref{sec:implementation}}. %\added{Note that one dimension is reduced due to azimuthal symmetry of the distribution about the orientation.}
In Fig.~\ref{fig:phaseVis}, we visualize two phase functions with different liquid IORs but the same particle IOR. The results show that surrounding the particle with liquid leads to increased forward scattering compared to when it is surrounded by air.

\paragraph{\textbf{Properties and relationship to the micro-flake model.}}
Our attenuation coefficients and phase function follows the framework by Jakob et al.~\shortcite{jakob2010radiative}, ensuring both the system reciprocity and the normalization of the phase function.

%While our micro-ellipsoid model and the micro-flake model have the same normal distribution function (ellipsoidal distribution), their main difference 
Our micro-ellipsoid model and the micro-flake modelis have different particle shape, where our particle is an ellipsoid, and the micro-flake relies on a flake without any volume. The advantage of the ellipsoid is that it allows for transmittability, although it lacks a closed-form formulation. The micro-flake model can also be considered as a special case of our model, by setting $\sigma_p=0$, which makes the ellipsoid into a disk.

%------------------------------------------------%
\subsection{Isotropic medium}
%------------------------------------------------%

In a special case, the particle shape become a sphere, by setting $\sigma_p=1$, leading to an isotropic medium. In this scenario, the volume extinction coefficients \(\sigma_t\) and \(\sigma_s\) become constants:
\begin{equation}
    \sigma_t =  \int_{\Omega} D(\omega_m) \, d\omega_m, \quad \sigma_s = \alpha \sigma_t.
\end{equation}

%In the special case of spherical particle shape ($\sigma_p=1$) where the projected area \(\sigma(\omega_m, \omega) = 1\), this model reduces to a traditional isotropic medium. 
The particle phase function and the micro-ellipsoid phase function are also simplified and only depends on \(\cos\theta = \omega_i \cdot \omega_o\):

%The micro-phase function now depends only on \(\cos\theta = \omega_i \cdot \omega_o\) and is uncorrelated with \(\omega_m\). Consequently, it can be factored out of the integral in Eqn.~(\ref{eq:micro_volume_sigma}). As a result, the phase function simplifies to the micro-phase function itself:
\begin{equation}
f_p(\omega_i, \omega_o) = p(\omega_i, \omega_o) \frac{\int_{\Omega} \sigma(\omega_m, \omega_i) D(\omega_m) \, d\omega_m}{\sigma_t(\omega_i)} = p(\omega_i, \omega_o).
\end{equation}

%\added{The isotropic phase function can be defined based on the angle $\theta$ between the incoming and outgoing directions, reducing it to one dimension. }
As the phase function only has one dimension, it can be easily represented by fitting a basis function. In practice, we use two Gaussians to represent the one-dimensional phase function: one for forward scattering and the other for backward scattering:
%To represent this, we propose a data-driven solution by fitting simulated data using a basis function. 
\begin{equation}
    \begin{aligned}
        f_p(\theta,\mathcal{S}) &= w_1 \cdot G_1(\theta; \mu_1,\sigma_1) + w_2 \cdot G_2(\theta; \mu_2,\sigma_2), \\
          G(x; \mu,\sigma) &= \frac{1}{\sqrt{2\pi}\sigma} e^{-\frac{(x-\mu)^2}{2\sigma^2}},
    \end{aligned}
\end{equation}
where $G_1$ and $G_2$ are two one-dimensional Gaussian functions with $\mu_1$ and $\mu_2$ as the mean values, $\sigma_1$ and $\sigma_2$ as the variances. $w_1$ and $w_2$ are the weights for two Gaussians. All these parameters are obtained by optimization using the Monte Carlo simulation as GT. 
\section{WetSpongeCake: A BSDF model with porosity and saturation }
%--------------------------%

%\myfigure{BSDFConfig}{bsdf.pdf}{ The light transport within a medium is approximated by a surface model, consisting of an analytical single scattering and a Monte Carlo-based multiple scattering by performing the integral along the depth dimension only rather than the entire 3D spatial domain. \added{do we still need this figure?} }

% \mycfigure{paper}{paper2.pdf}{ A wet paper with different saturation textures (bottom right). Our method is able to characterize these different kinds of appearances with a low time cost (about 34 seconds) using physical materials, while no existing approaches can achieve this. Medium parameters: $T=0.0017, n=300M, P=0.5, \sigma_t^l = 0$. Particle parameters: $\alpha=(1,1,1), \eta_l=1.333$, 70\% cellulose + 30\% $\text{TiO}_2$.}

We have already defined the medium, which still requires large sampling rates for convergence. To address this, we integrate it into the SpongeCake framework by treating the medium as a layer. We derive an analytical model for single scattering, taking into account both reflection and transmission. Then, we further introduce multiple scattering, delta transmission, and the air-liquid interface. 
%\added{what's the challenge to do this.}

\subsection{Single scattering}
The single scattering within a medium with thickness $Z$ can be computed by the integral over the depth of the single scattering vertex~\cite{Wang:2022:sponge}. At each single scattering vertex, we compute phase function, the extinction of the particles and the liquid, together with cosine terms due to the change of the integration domain. Then, the BRDF and BTDF can be formulated as: 
\begin{equation}
\begin{aligned}
f_{r}\left(\omega_{i}, \omega_{o}\right)=&\int_{0}^{Z}\frac{\alpha\sigma_t(\omega_i) f_{p}\left(\omega_{i},\omega_{o}\right)}{\cos \omega_{i}  \cos \omega_{o}} T\left(\frac{t}{\cos\omega_i}\right)T\left(\frac{t}{\cos\omega_o}\right) \mathrm{d}t \\
=& K \sigma_s(\omega_i) f_p\left(\omega_{i},\omega_{o}\right)  \frac{1-e^{-\left(\sigma_i\frac{Z}{\cos\omega_i}+\sigma_o\frac{Z}{\cos\omega_o}\right) }}{\sigma_i \cos\omega_o + \sigma_o \cos\omega_i}, \\
f_{t}\left(\omega_{i}, \omega_{o}\right)=&\int_{0}^{Z}\frac{\alpha\sigma_t(\omega_i) f_{p}\left(\omega_{i},\omega_{o}\right)}{|\cos \omega_{i}|  |\cos \omega_{o}|} T\left(\frac{t}{|\cos\omega_i|}\right)T\left(\frac{Z-t}{|\cos\omega_o|}\right) \mathrm{d}t\\
=&K \sigma_s(\omega_i) f_p\left(\omega_{i},\omega_{o}\right)  \frac{e^{-\frac{\sigma_i Z}{|\cos\omega_i|}}-e^{-\frac{\sigma_o Z}{|\cos\omega_o|}}}{\sigma_o|\cos\omega_i|-\sigma_i|\cos\omega_o|},
\end{aligned}
\end{equation}
where $\sigma_i=K\sigma_t(\omega_i)+S\sigma_t^l$ and $\sigma_o=K\sigma_t(\omega_o)+S\sigma_t^l$. The single scattering can be computed using the precomputed phase function.

\subsection{Multiple scattering}
%Multiple scattering plays a significant role in materials with high albedo or large thickness. 
%However, there is no analytical solution available for multiple scattering. While a Monte Carlo random walk can provide unbiased results, it is computationally expensive and introduces additional variance. 
To compute multiple scattering efficiently, Wang et al.~\shortcite{Wang:2022:sponge} estimate the multiple scattering of a BSDF by mapping it into a single-scattering lobe with modified parameters along with a Lambertian term with a simple neural network. Their approach could be applied to our BSDF as well. However, the key challenge is that our phase function is not analytical but represented by a precomputed table, which lacks differentiability. To address this limitation, we represent phase function using a neural network to enable multiple to single scattering mapping.

Specifically, we propose a phase function network and a parameter mapping network. The former transforms phase function parameters into a phase function value using a simple multi-layer perceptron (MLP). The parameter mapping network converts multiple scattering parameters into single scattering parameters, together with a Lambertian parameter and their weights. The phase function neural network is trained first. Afterward, the mapping network is trained by rendering the predicted single scattering parameters with the learned phase function network. Once the mapping network is trained, we tabulate the phase function to avoid network inference during rendering. More details are in the supplementary.

Please note that we do not utilize the neural representation of the phase function for the single scattering, as it is much sharper, which is beyond the capabilities of the learned network.

In the case of an isotropic medium, since the phase function can be represented using two Gaussian distributions, we can directly learn the mapping network without the need to train an additional phase function neural network.

\subsection{The other components}

\paragraph{\textbf{Delta transmission.} }
For a thin medium, light can pass through without any scattering, such as paper and cloth. This delta function can be determined by analyzing the transmission of light through the medium in the direction of $\omega_i$, similar to the SpongeCake model.

\paragraph{\textbf{Air-liquid interface. }}
When a material is fully saturated ($S=1$), a thin film of liquid forms on the surface. Our model captures the effects of light reflection and refraction at the interface by layering a dielectric BSDF on top of the medium.

\paragraph{\textbf{Importance sampling.}}
Importance sampling is a necessary component for a BSDF. For this, we use a simple solution, by sampling the phase function with cumulative distribution function to get the outgoing direction.

\section{Results}
\label{sec:results}

We have implemented our algorithm inside the Mitsuba renderer \shortcite{Mitsuba}. 
% We also implement our modified RTE, which considers the porosity and saturation as the ground truth (GT), since no other public code is available for wet poroused materials. 
% We use mean square error (MSE) to measure the difference between each method and our modified RTE. 
All timings in this section are measured on a 2.20GHz Intel i7 (48 cores) with 32 GB of main memory. We provide the material settings in the supplementary material.

\subsection{Model validation}
% \added{please say some there here.}
We validate our model through a series of experiments, including comparisons against photographs, equal-time comparison between the surface model and volume model, validation of multiple scattering, and the white furnace test. Please note that our multiple scattering model is applied to all materials, except for the spatially varying anisotropic material, for which we perform a random walk in a position-free framework~\cite{Guo:2018:Layered}.% \added{as a table for each parameter within the texture is infeasible.}

\begin{figure}[h]
\centering
\includegraphics[width = \linewidth]{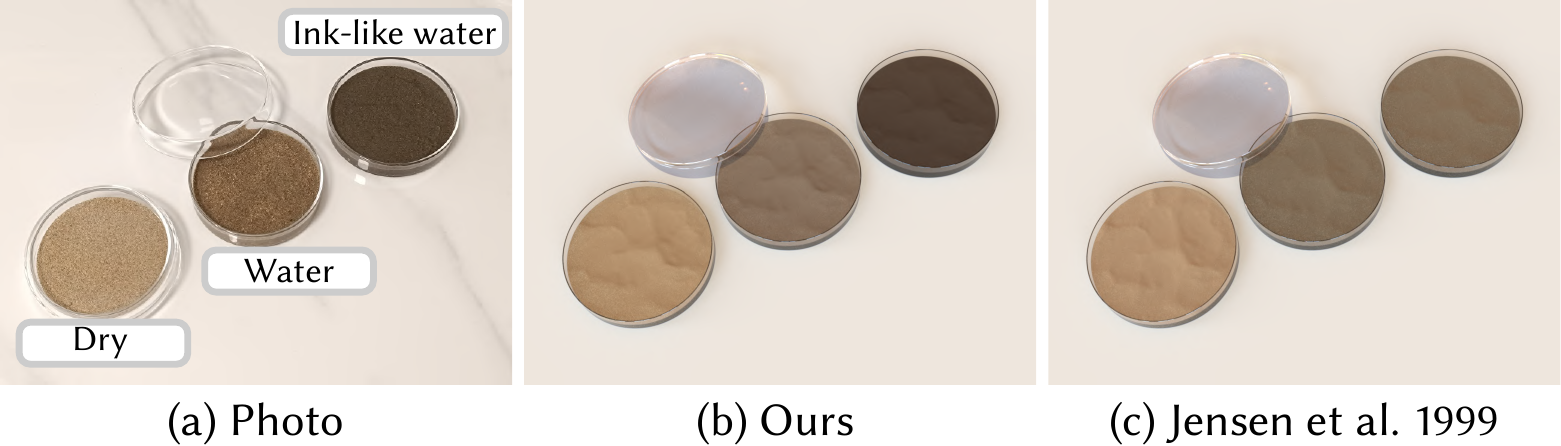}
\caption{Comparison between the photograph, our model, and the two-HG method by Jensen et al.~\shortcite{Jensen1999}. Our model demonstrates a close match with the photograph, whereas the two-HG method by Jensen et al.~\shortcite{Jensen1999} captures the change in scattering direction but fails to adequately model liquid absorption. }
\label{fig:sand}
\end{figure}

\paragraph{Sand scene}
To demonstrate WetSpongeCake's capability to replicate natural appearances, we compare our model, Jensen et al.~\shortcite{Jensen1999}, which employs two Henyey-Greenstein (HG) lobes as the phase function, along with a real photograph. The comparison is shown in Fig.~\ref{fig:sand}. This scene consists of three types of sand (isotropic) in different containers: dry, saturated with water, and saturated with ink, respectively, lit under an environment map and a directional light. We render this scene with both single and multiple scattering, considering global illumination. Our model almost matches the photograp, while Jensen et al.'s method, which adjusts HG parameters manually, can not model liquid absorption.

\begin{figure}[h]
\centering
\includegraphics[width = \linewidth]{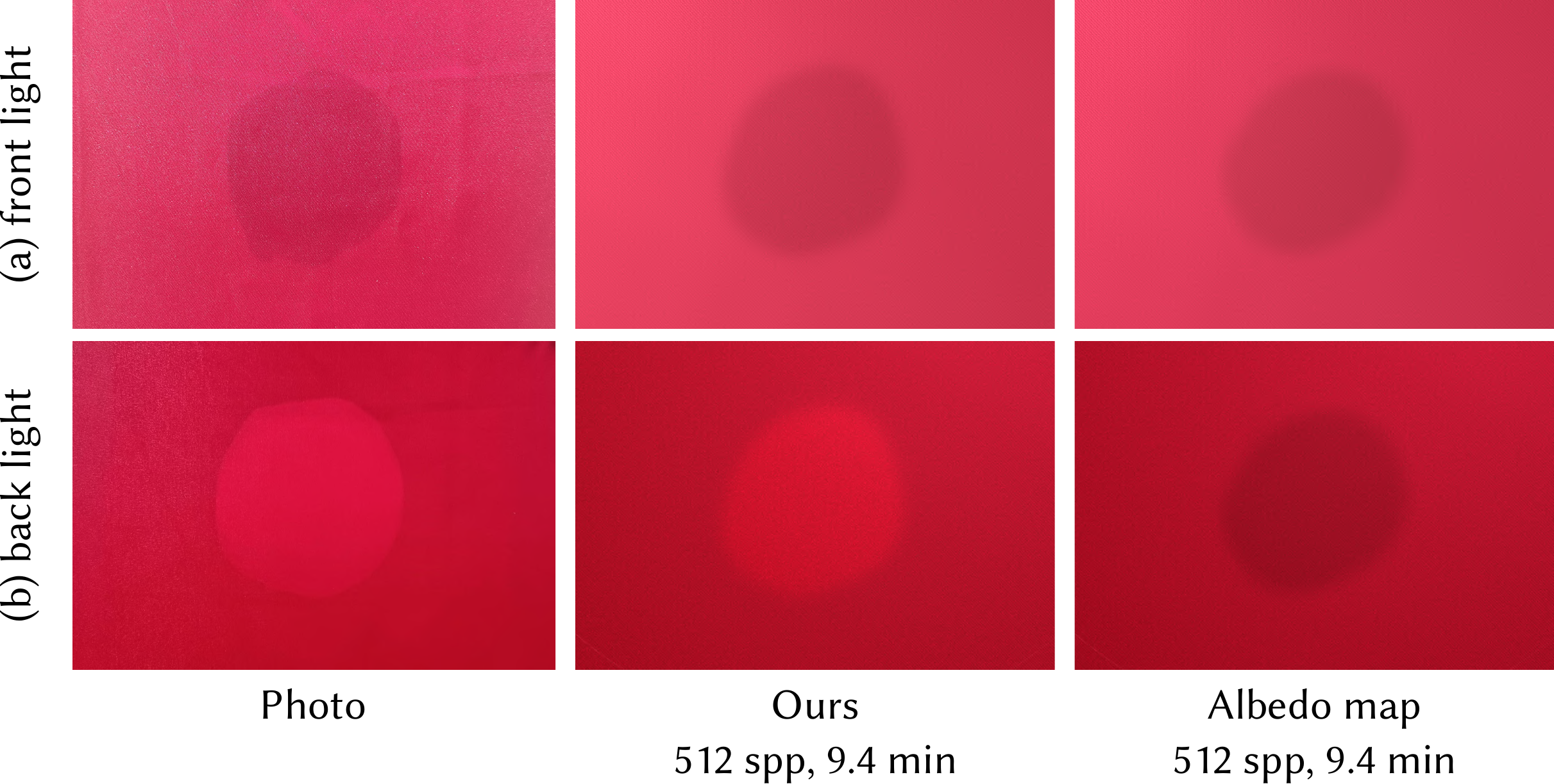}
\caption{Comparison between photographs, our model, and an empirical approach which modifies the albedo at two light conditions (front-lit and back-lit). In the photographs, the wet regions of the cloth exhibit reduced reflection and enhanced transmission. Our model replicates these effects accurately, whereas using a darker albedo achieves a similar reduction in reflection but results in incorrect darkening of transmission.}
\label{fig:cloth}
\end{figure}

\paragraph{Cloth scene}
In Fig.~\ref{fig:cloth}, we compare our method, the albedo map method, which achieves darker reflections by adjusting the albedo, along with a real photograph. The scene consists of a wet cloth (anisotropic medium) under two lighting conditions: front-lit and back-lit. We render this scene with both single and multiple scattering. The photographs demonstrate that the wet regions of the cloth exhibit reduced reflection and enhanced transmission, a behavior accurately captured by our model. In contrast, simply modifying the albedo reduces both reflection and transmission, failing to reproduce the desired appearance.

\begin{figure}[h]
\centering
\includegraphics[width = 0.9\linewidth]{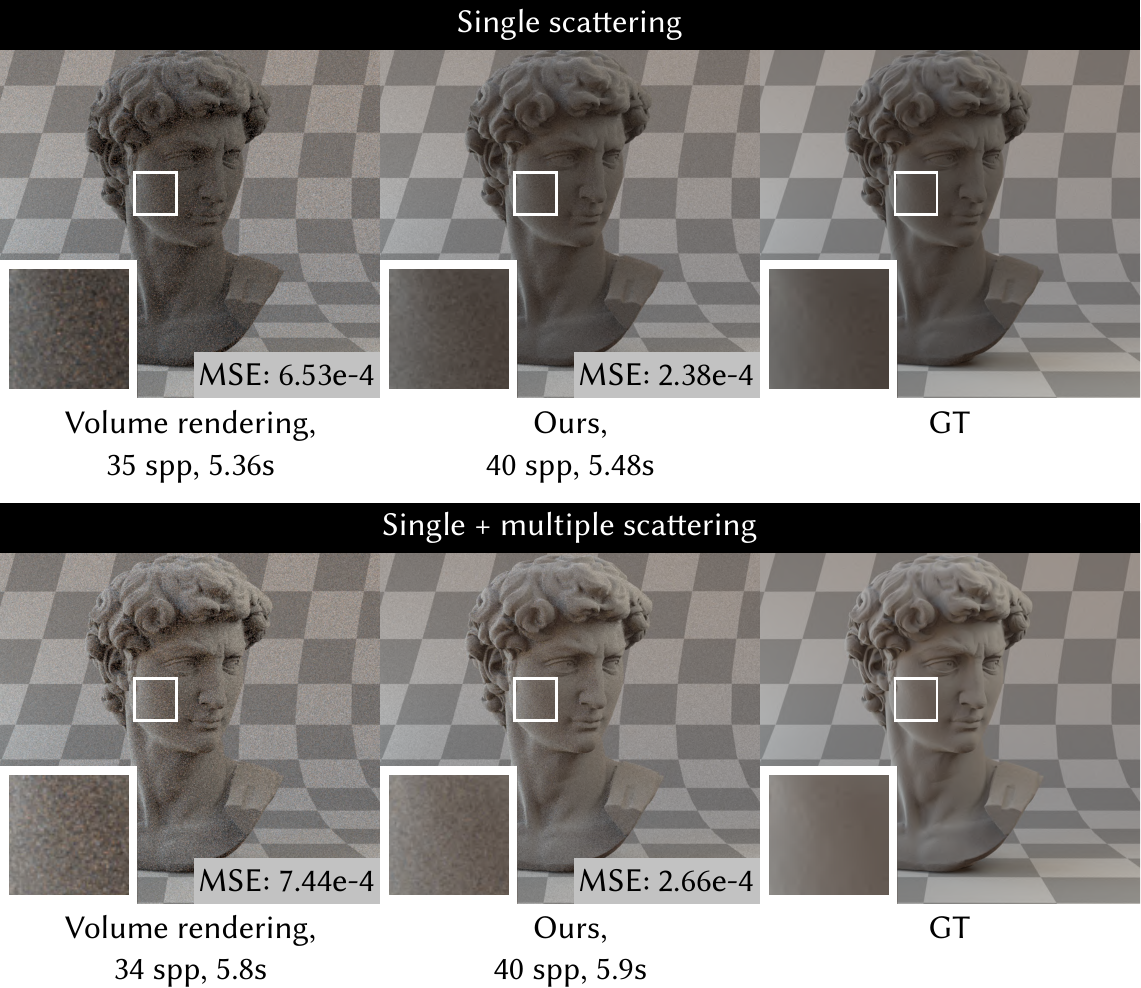}
\caption{Comparison between our BSDF surface model and the volume rendering with our modified RTE. The references are rendered with our modified RTE at a high sample rate. In both cases, our BSDF produces results with much less noise and lower error than the volume renderings with equal time.}
\label{fig:david_head}
\end{figure}

\paragraph{Our volume model vs. our surface model.}
We validate the effectiveness of our surface model (both single and multiple scattering), by comparing it to our volume model in Fig.~\ref{fig:david_head}, where the converged volume renderings are treated as GT. We find that the result of our single scattering is almost identical to the GT while having much less noise than the volume rendering with equal time, as it is analytical and does not need a random walk. After introducing the multiple scattering, the result of our surface model still exhibits less noise and lower error than volume rendering with equal time, thanks to our effective way of computing the multiple scattering. 

%In this scene, we consider direct illumination under two area light sources and an environment map. We compare our method against the volume rendering (our modified RTE), where the converged volume rendering result is treated as the ground truth.

% Thanks to the position-free property, although our method needs Monte Carlo simulation for multiple scattering, the integration is performed on the depth rather than the entire spatial domain, significantly reducing noise.

\begin{figure*}[h]
\centering
\includegraphics[width = 0.9\linewidth]{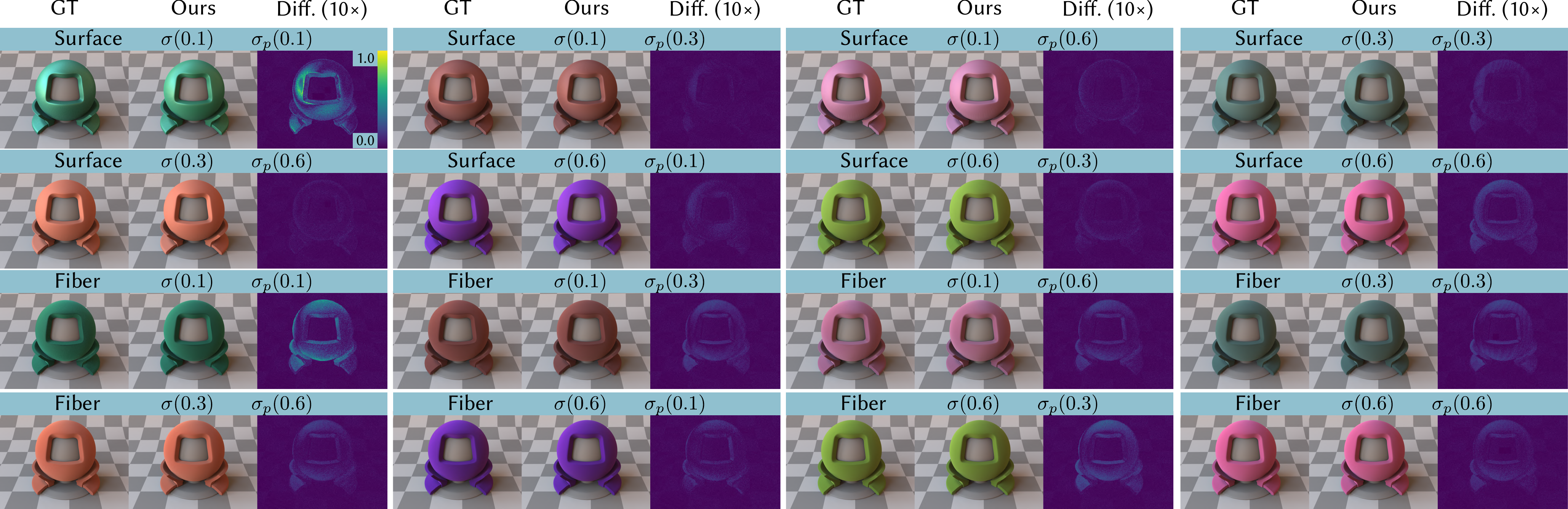}
\caption{Multiple scattering validation. For each example, we provide the normal projected area $\sigma$, particle projected area $\sigma_p$. The ground truth is obtained through Monte Carlo simulation of multiple scattering. Our results exhibit minor differences from the ground truth in most cases.}
\label{fig:multipleScattering}
\end{figure*}

\begin{figure*}[h]
\centering
\includegraphics[width = 0.9\linewidth]{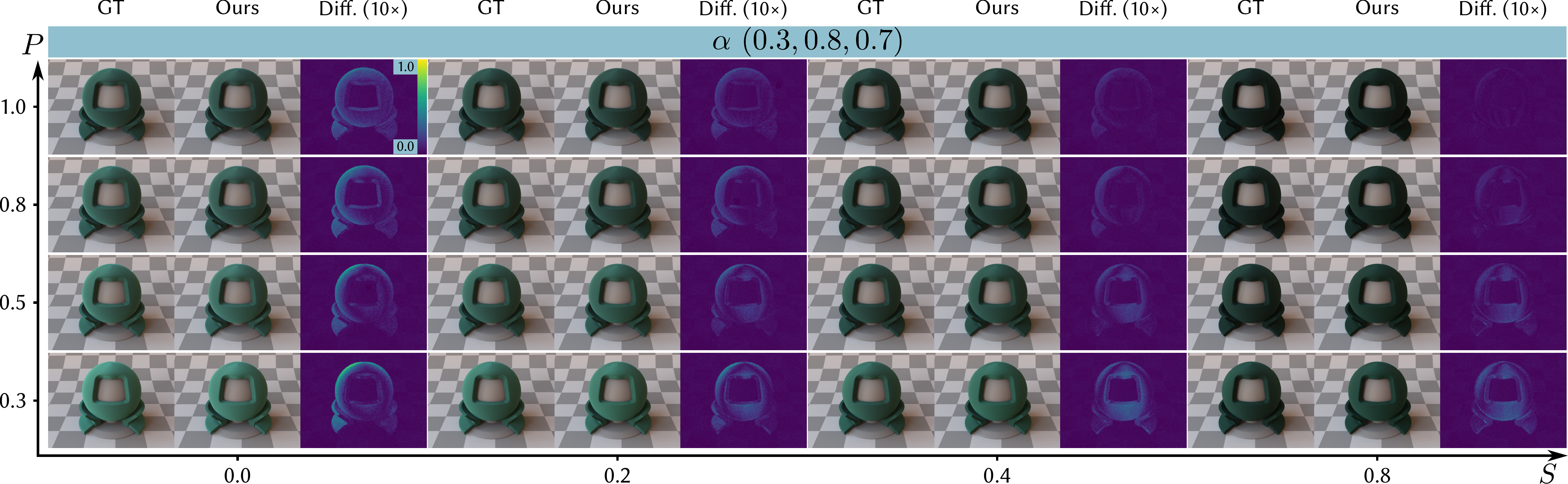}
\caption{Multiple scattering validation for a set of materials, with varying porosity $P$ and saturation $S$. Monte Carlo simulation of multiple scattering is used as the ground truth. Our results show only minor deviations from the ground truth.}
\label{fig:multipleScattering_2}
\end{figure*}

\paragraph{Multiple scattering validation}
In Figs. \ref{fig:multipleScattering} and \ref{fig:multipleScattering_2}, we validate our multiple scattering by comparing its results to the GT obtained by Monte Carlo random walk~\cite{Guo:2018:Layered} across various materials. The results demonstrate an overall good match, except in cases with a low projected area where the sharp distribution cannot be accurately expressed by the phase function network.
%compare the rendering results and scattering lobes of our multiple scattering model with the ground truth obtained via Monte Carlo random walk across various materials. 

%The Monte Carlo random walk used for the ground truth follows a similar approach to Guo et al.~\shortcite{Guo:2018:Layered}, but without considering surface interfaces. 

\begin{figure}[h]
\centering
\includegraphics[width = 1\linewidth]{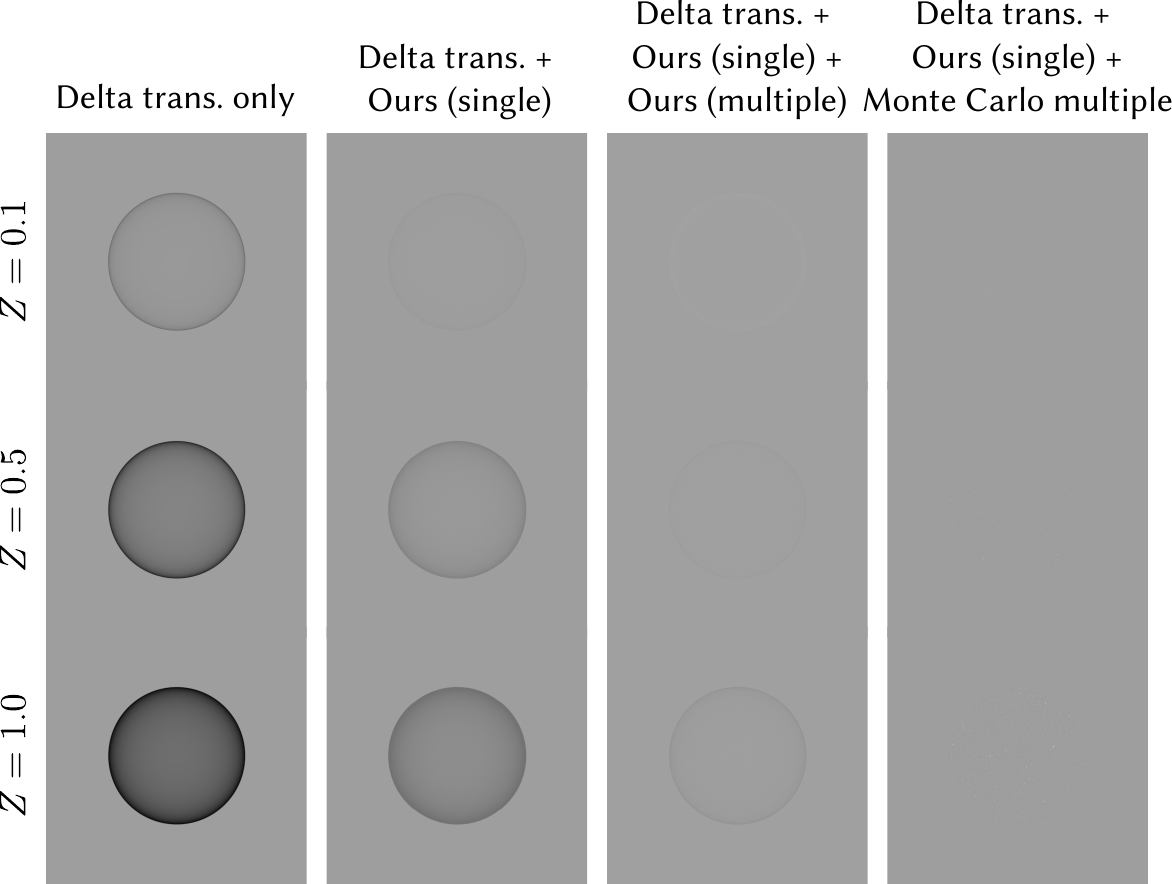}
\caption{A white furnace test for a WetSpongeCake material with no absorption. While the delta transmission and single scattering result in darker pixel values, our multiple scattering model produces an image that is nearly uniform, with minor inaccuracies. The Monte Carlo multiple scattering successfully passes the white furnace test.}
\label{fig:white_furnace}
\end{figure}

\paragraph{White furnace test}
We perform a white furnace test for our surface model on materials with no absorption, under a constant environment map with radiance set as one. As shown in Fig.~\ref{fig:white_furnace}, the delta transmission and the single scattering results in darker pixel values, as expected. When including the multiple scattering using Monte Carlo random walk, a constant image was produced, while our multiple scattering closely approximates a constant image, although there are some inaccuracies due to network bias.

\begin{figure*}[h]
\centering
\includegraphics[width = 0.7\linewidth]{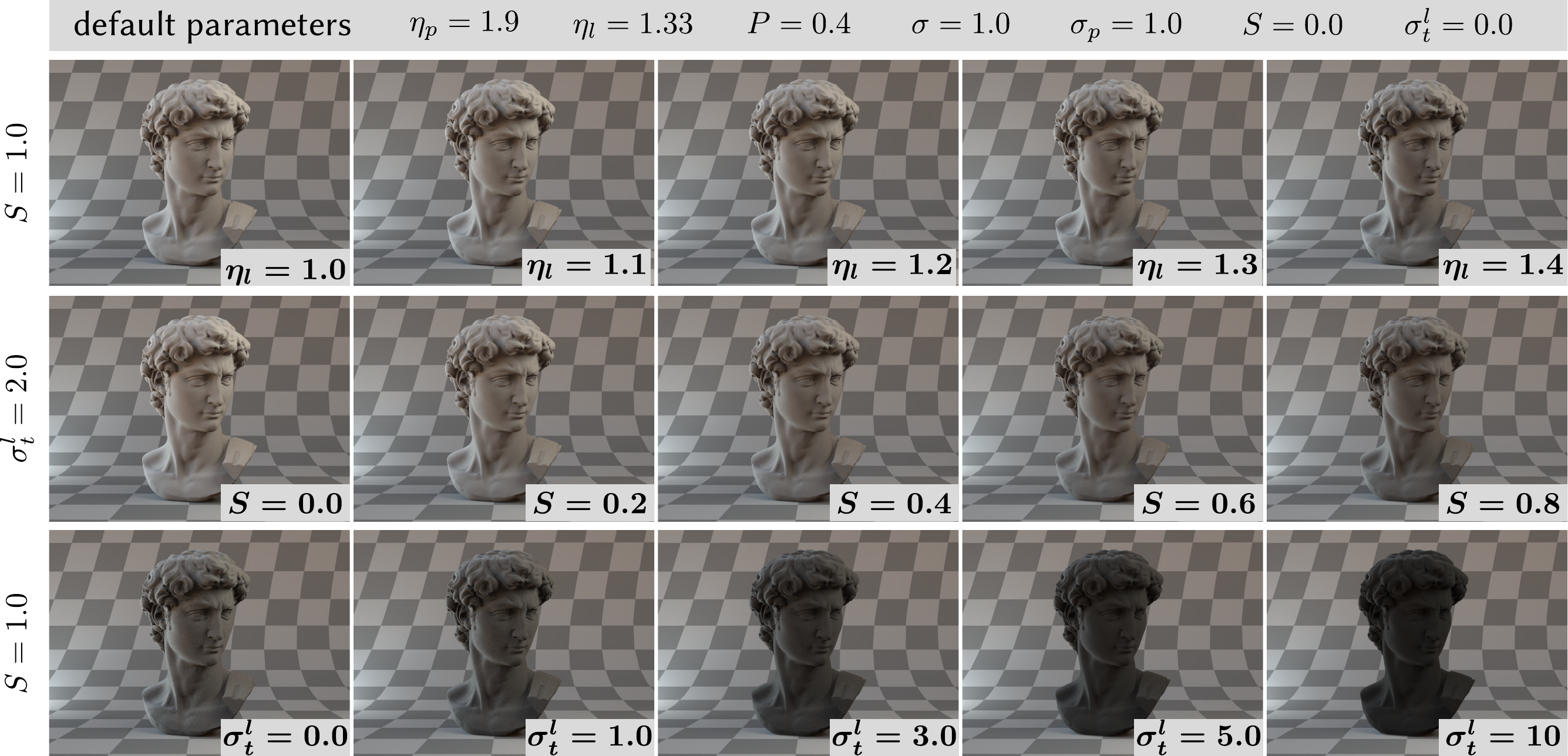}
\caption{Sculpture scene under different parameter settings. Parameters not explicitly mentioned are set to their default values, as indicated at the top of the figure.}
\label{fig:parameters}
\end{figure*}

\paragraph{Sculpture Scene}  
To demonstrate the impact of different parameters on appearance, we present renderings of the Sculpture scene under various parameter settings in Fig.~\ref{fig:parameters}. This scene is illuminated by two area light sources and an environment map. As the liquid's IOR increases, the phase function becomes more forward-scattering, causing a darker appearance. Similarly, higher saturation enhances forward scattering while incorporating liquid absorption, further darkening the sculpture. Furthermore, increased light absorption by the liquid results in a noticeably darker sculpture.

\subsection{More results}

% \paragraph{Phase function validation}
% Our two-Gaussian phase function is a critical component of our method. We validate its effectiveness in Fig.~\ref{fig:phaseRendering} and Fig.~\ref{fig:phaseCurve2} by comparing the rendering results and the curves of several different basis functions on several types of particles. We compare our two-Gaussian functions, one Gaussian, two HG functions, and the simulated data (performing Monte Carlo for a particle), which we treat as reference. A single-Gaussian function shows the worst match with the simulated data, due to its limited capability. The two-HG function can represent the backward scattering due to the second lobe, but its distribution differs from the simulated data, even for the forward scattering. Our two-Gaussian function can match the GT closely and consistently on all shown particles. 

% \paragraph{Sculpture scene}
% To demonstrate the appearance differences between our model and other models,
% we present the rendering results of four models in Fig.~\ref{fig:4models}, including Lambertian, Lambert-sphere~\cite{d2021analytic}, Chandrasekhar's BRDF, and ours. For ease of comparison, we use isotropic phase function in our model. Note that if we use a different phase function, our model will show a totally different appearance. When porosity approaches 1, our model exhibits the same appearance as Chandrasekhar's BRDF. As porosity decreases, the appearance of our model becomes brighter.

\paragraph{Teaser scene}
In Fig.~\ref{fig:teaser}, we show several objects with porous materials lit by an environment map with global illumination, including wet cloths (anisotropic), flowerpots (isotropic), and sculptures (isotropic). Our BSDF can represent a wide range of appearances in the real world with high fidelity, controlled by physical parameters. %It is simple and practical, so we believe it can serve as a plugin for existing renderers. 

%\added{We use the network solution for multiple scattering of flowerpots, while random walk is applied for the cloths.} 

\begin{figure}[h]
\centering
\includegraphics[width = 1.0\linewidth]{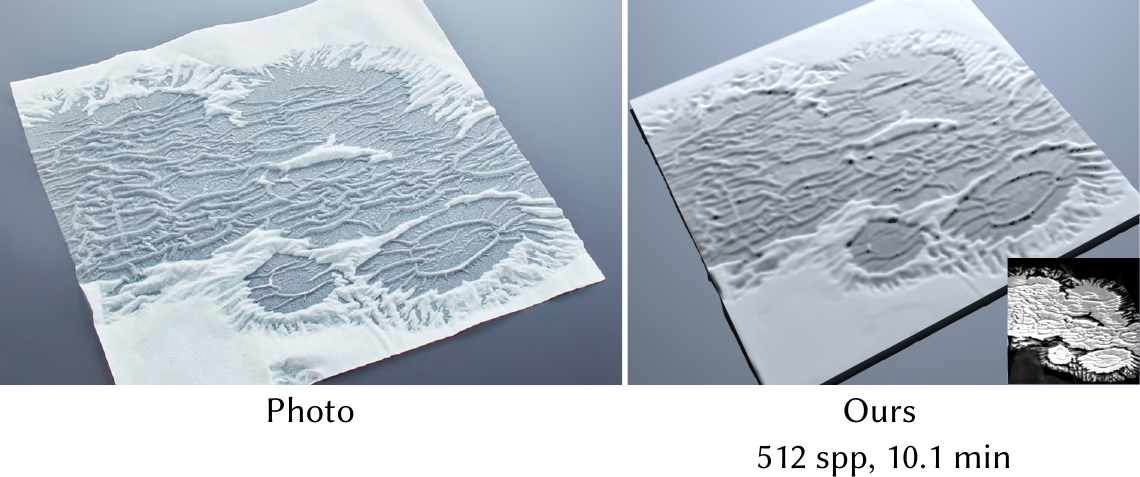}
\caption{A wet paper saturated by water. Our method is able to characterize the saturation level using a saturation map.}
\label{fig:paper}
\end{figure}

\paragraph{Paper scene}  
We design a Paper scene using the photo converted to a grayscale image as a saturation map, as shown in Fig.~\ref{fig:paper}. The saturation map, displayed in the bottom-right corner of the rendered image, defines the varying saturation across the surface. This scene is illuminated by an environment map, taking global illumination into account. Our model is able to produce the desired effects.

\begin{figure}[h]
\centering
\includegraphics[width = 1\linewidth]{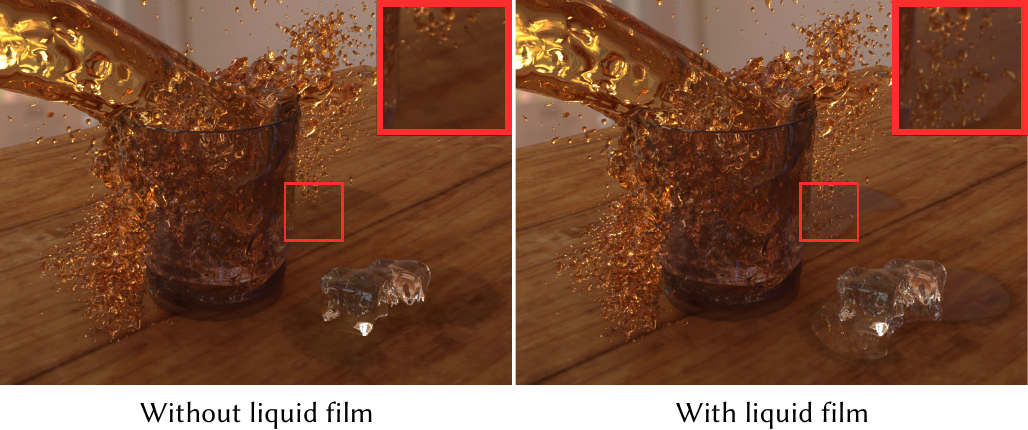}
\caption{Our model supports a thin liquid film on top of the medium, capturing the reflection induced by the air-liquid interface.}
\label{fig:Table}
\end{figure}

\paragraph{Table scene}
In Fig.~\ref{fig:Table}, we present the appearance effects caused by the air-liquid interface. This scene consists of wood (isotropic), illuminated by an environment map, with global illumination taken into account. The table surface without a liquid film appears diffuse, while the table surface with a liquid film exhibits a specular lobe.

% \paragraph{Sand scene}
% In this Sand scene (Fig.~\ref{fig:sand}), we show both dry and wet sand, which are filled with water or benzene. The sand consists of three particles (40\% $\text{Fe}_2\text{O}_3$ + 40\% $\text{FeO(OH)}$ + 20\% $\text{SiO}_2$) with a spatially-varying albedo, the porosity as 0.425, and the saturation as 0.9. The other parameters for benzene and water are $ n_l=1.501, \sigma_t^l = 0$ and $n_l=1.333, \sigma_t^l = 0$, respectively. We render this scene with both single and multiple scattering under an environment map with global illumination. Our model is able to provide vivid and reasonable results. With the same saturation, benzene should give a darker appearance than the water due to their different refraction indices. Similarly, wet sand looks darker than dry sand. Our model can well characterize both of these effects. 

% \paragraph{Phase function visualization}
% We present a visualization of the phase function of the micro-volume model along with its corresponding mean cosine across different parameter settings. Fig.\ref{fig:phaseCurve_iso} illustrates the isotropic case, while Figs.\ref{fig:phaseCurve_ani_eta} and \ref{fig:phaseCurve_ani_sigma} illustrate anisotropic cases. A phase function with a high $\eta_p$ exhibits a larger mean cosine, indicating stronger backward scattering. In contrast, a high $\eta_l$ results in a smaller mean cosine, corresponding to enhanced forward scattering, as we expected.

\subsection{Discussion and limitations}
We have identified several limitations in our method. Our model does not have closed-form formulations and relies on precomputation, which restricts its ability to represent multiple scattering in spatially varying anisotropic materials. Additionally, similar to previous studies~\cite{Wang:2022:sponge}, our multiple scattering network may sometimes struggle to achieve accurate fits.

\section{Conclusion}
In this paper, we have presented a transmittable micro-ellipsoid model for rendering wet porous materials, applicable for both volume and surface appearance rendering. We generalize the anisotropic radiative transfer equation to account for both porosity and saturation, which enables volume rendering of wet porous materials together with the micro-ellipsoid model. Building on this foundation, we then derive a practical surface appearance model under the position-free framework, termed WetSpongeCake. The WetSpongeCake model effectively represents the diverse appearances of wet porous materials using physically meaningful parameters such as porosity and saturation, and it captures both reflection and transmission effects. The resulting appearances closely mimic real-world observations. To the best of our knowledge, no existing method offers such a lightweight, physically-based shading model for wet porous materials. Finding a closed-form solution for the phase function is crucial for completing the theory and further enhancing the model's practical applicability. We also believe that our model has potential applications in other areas, such as inverse rendering.

\clearpage  
\bibliographystyle{ACM-Reference-Format}
\bibliography{paper}

\clearpage 

% \mycfigure{phase_curve}{phaseCurve.pdf}{Comparison among two Gaussians (ours), one Gaussian, two HG functions, and the simulated data (GT) on different types of particles and liquid. Our two-Gaussian phase function has the closest match with the GT on all these settings compared to other basis functions.}

% \mycfigure{single_validation}{single_validation.pdf}{Single validation. Equal spp (1), equal time (about 0.13s). GT generated from medium, using 1024 spp. 没有使用matpreview而是使用了sphere，因为几何会有问题（使用了medium的几何会有明显的面片，而使用了BSDF的几何会对法线进行插值）}

% \mycfigure{single_multiple}{single_multiple.pdf}{和前面一样的原因，reference使用mesh可能会渲染出明显的面片，暂时选择sphere来展示}

%\mycfigure{whitefurnace}{whitefurnace.pdf}{\added{Whitefurnace test.}}

\end{document}


\title{Supplemental materials: A Micro-Ellipsoid Model for Wet Porous Materials Rendering}

% \title{Height-free Multiple-bounce Smith Microfacet BRDFs}
%multiple-bounce, path space.
% Shadowing-masking

\author{Gaole Pan}
\orcid{0009-0007-9335-333X}
\affiliation{%
  \institution{Nanjing University of Science and Technology}
  %\city{Nanjing}
  \country{China}
}
\email{pangaole@njust.edu.cn}

\author{Yuang Cui}
\orcid{0009-0006-8983-7844}
\authornote{Research done when Yuang Cui was an intern at Nanjing University of Science and
Technology. }
\affiliation{%
  \institution{Anhui Science and Technology University}
  %\city{Bengbu}
  \country{China}
}
\email{yuangcui@outlook.com}

\author{Jian Yang}
\orcid{0000-0003-4800-832X}
\affiliation{%
  \institution{Nanjing University of Science and Technology}
  %\city{Nanjing}
  \country{China}
}
\email{csjyang@njust.edu.cn}

\author{Beibei Wang}
\orcid{0000-0001-8943-8364}
\authornote{Corresponding author.}
%\thanks{$^\dagger$Corresponding author.}
\affiliation{
    \institution{Nanjing University}
    \country{China}
}
\email{beibei.wang@nju.edu.cn}
%\affiliation{%
%  \institution{Nankai University}
%  \city{Tianjin}
%  \country{China}
%}
%\affiliation{%
%  \institution{Nanjing University of Science and Technology}
%  \city{Nanjing}
%  \country{China}
%}

\renewcommand{\shortauthors}{Pan et al.}

\maketitle

% \input{sup_sec_analysis}
% % \input{pseudocode}
\section{Implementation details}
\label{sec:implementation}
In this section, we present additional details about our implementation, including the simulation process for the particle phase function and the precomputation of the attenuation coefficients and phase function (Sec.~\ref{sec:micro-ellipsoid_impl}). For WetSpongeCake, we provide further details on the neural network (Sec.~\ref{sec:wetspongecake_impl}).

\subsection{Micro-ellipsoid model}
\label{sec:micro-ellipsoid_impl}
\paragraph{particle phase function}

% \added{The particle phase function is challenging to derive analytically due to the complexity introduced by multiple scattering events within the particle. Similarly, evaluating the extinction coefficient and phase function involves solving the integrals in Eqn.~(\ref{eq:micro_volume_sigma}) and Eqn.~(\ref{eq:micro_volume_phase}), which lack closed-form analytical solutions. To overcome these challenges, we adopt a precomputation strategy. This approach allows us to tabulate the required values in advance, enabling efficient lookup and usage during runtime.}

The particle phase function is computed via Monte Carlo simulation and stored as a 3D lookup table. The center of the ellipsoid is positioned at the origin of the coordinate system, and the particle orientation \(\omega_m\) aligns with the z-axis. For each incoming direction \(\omega_i\), we generate the starting point of the ray by sampling a disk located in the plane perpendicular to the incoming ray, ensuring that it can cover the entire ellipsoid. We then trace the ray and check for intersections with the ellipsoid. If no intersection occurs, the ray is discarded. If an intersection is found, we perform regular path tracing, where at each intersection, we decide whether the ray is reflected or refracted based on the Fresnel term as the probability. The process continues until the ray exits the particle.

We store the table with the following dimensions: the polar angle $\theta_i$ of \(\omega_i\), the polar angle $\theta_o$ of \(\omega_o\), and the azimuthal angle $\phi_o$ of \(\omega_o\). Due to the azimuthal symmetry of the particle shape about the normal $m$, the azimuthal angle of the incident direction can be omitted. 

\paragraph{Attenuation coefficients and phase function}
The attenuation coefficients and the phase function of the aggregated medium are computed using Monte Carlo estimation:
\begin{equation}
    \begin{aligned}
        \sigma_t(\omega) &= \frac{1}{N}\sum_{i=0}^{N-1}\frac{\sigma(\omega_m^i, \omega) D(\omega_m^i)}{\mathrm{pdf}(\omega_m^i)},\\
        f_p(\omega_i,\omega_o) &= \frac{1}{N}\sum_{i=0}^{N-1}\frac{p(\omega_m^i, \omega_i , \omega_o) \sigma(\omega_m^i, \omega_i) D(\omega_m^i)}{\sigma_t(\omega_i) \mathrm{pdf}(\omega_m^i)},
    \end{aligned}
\end{equation}
where $N$ is the number of samples, and $\mathrm{pdf}(\omega_m^i)$ is the probability density function (PDF) for the $i$-th sample. Specifically, the attenuation coefficients are stored as a 1D lookup table, parameterized by the polar angle $\theta$ of $\omega$, while the phase function is stored as a 3D lookup table, with dimensions corresponding to the polar angle $\theta_i$ of $\omega_i$, the polar angle $\theta_o$ of $\omega_o$, and the azimuthal angle $\phi_o$ of $\omega_o$, similar to the particle phase function table. In practice, we importance sample the normal distribution to improve the estimation efficiency.

% However, we can achieve an unbiased evaluation at the cost of precomputation. Specifically, we tabulate the extinction coefficient and phase function by Monte Carlo estimation:
% \begin{equation}
%     \begin{aligned}
%         \sigma_t(\omega_i) &= \frac{1}{N}\sum_{i=0}^{N-1}\frac{\sigma(\omega_m^i, \omega_i) D(\omega_m^i)}{\mathrm{pdf}(\omega_m^i)},\\
%         f_p(\omega_i,\omega_o) &= \frac{1}{N}\sum_{i=0}^{N-1}\frac{p(\omega_m^i, \omega_i , \omega_o) \sigma(\omega_m^i, \omega_i) D(\omega_m^i)}{\sigma_t(\omega_i) \mathrm{pdf}(\omega_m^i)},
%     \end{aligned}
% \end{equation}
% where $N$ is the number of samples, and $\mathrm{pdf}(\omega_m^i)$ is the probability density function (PDF) for the $i$-th sample. In practice, we can importance sample the normal distribution $D$ to improve the efficiency.

\subsection{WetSpongeCake}
\label{sec:wetspongecake_impl}

\begin{figure}[t]
\centering
\includegraphics[width = 1\linewidth]{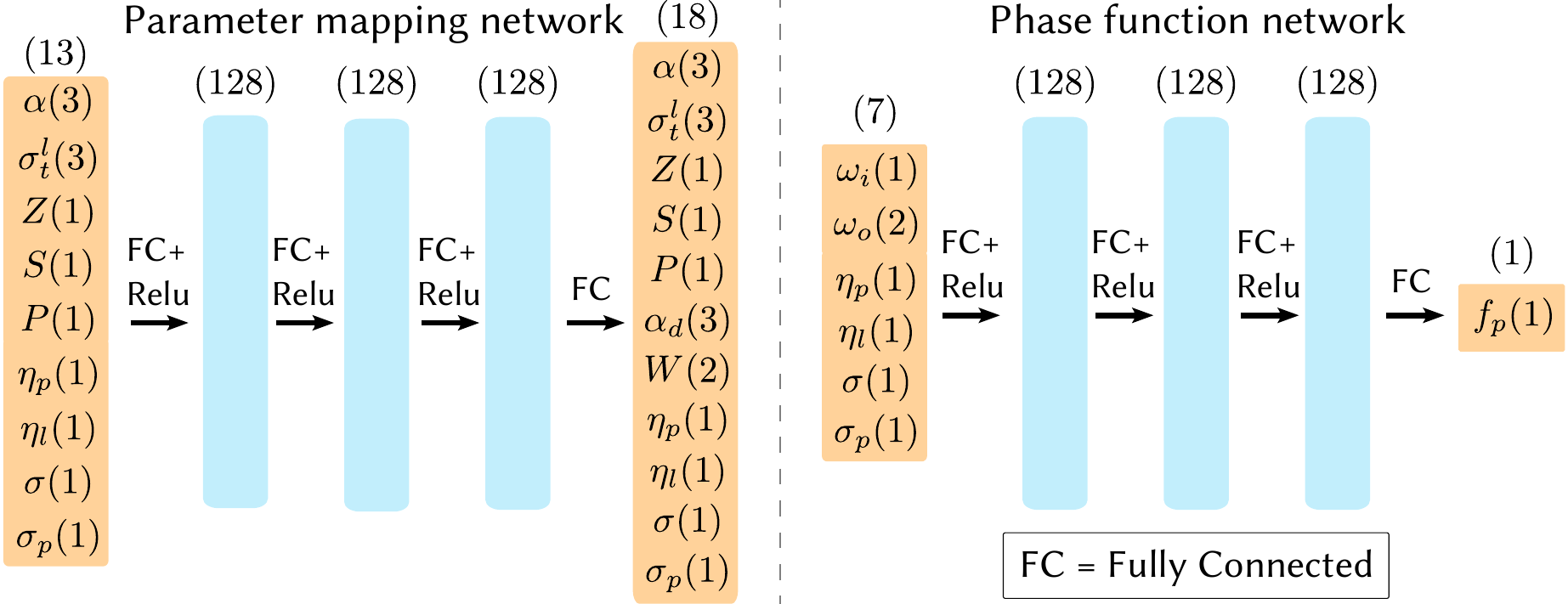}
\caption{The structures of two neural networks: the parameter mapping network (left) and the phase function network (right).}
\label{fig:net}
\end{figure}

\paragraph{Neural network structure}
The phase function network is a multi-layer perceptron (MLP) with three hidden layers, each containing 128 neurons, as shown in Fig.~\ref{fig:net}. Its input comprises phase function parameters and query directions, while its output is the phase function value. The parameter mapping network shares the same structure as the phase function network but differs in its input and output. Its input includes multiple scattering parameters. The output provides parameters for equivalent single scattering term, albedo $\alpha_d$ for the Lambertian term and the weights $W$ for the two term. 

\paragraph{Dataset}
For the phase function network, we generate 6,000 phase functions by randomly sampling input parameters. Directions $\omega_i$ and $\omega_o$ are sampled using $64 \times 64$ uniform stratified samples. We use $90\%$ of the dataset for training and reserve the remaining $10\%$ for validation. 

For the parameter mapping network, we generate 4,000 BSDFs by randomly sampling input parameters to create the training dataset. For each BSDF, we sample $\omega_i$ and $\omega_o$ using 32 × 32 uniform stratified samples. Starting from a sampled $\omega_i$, Monte Carlo sampling is performed in the medium by repeatedly selecting a new position and a new direction based on the phase function. This process continues until the maximum depth of 10 is reached or the ray exits the surface. For each $\omega_i$, we trace 10,000 rays. Although the data contains noticeable noise, it does not affect the neural network training. The remaining configuration is consistent with the phase function network. 

\paragraph{Training}
The loss function for both two networks is the mean absolute error (MAE) between the ground truth and the network's output. Both the neural network and our single scattering model are implemented in the PyTorch framework, enabling automatic differentiation. We use the Adam 
optimizer with a learning rate of 0.001. The network is trained with a mini-batch size of 16. Phase function network took 1 hour and parameter mapping network took 4 hours training on an NVIDIA 3090 GPU .

\section{Detailed Derivation of Parameter $K$}
\label{sec:derivationK}

In this section, we present the derivation of $K$ as outlined by Hapke \shortcite{Hapke2008}.

Hapke defined a porous medium where the porosity $P$ represents the fraction of the volume not occupied by solid particles, $n$ denotes the total number of particles per unit volume, and $E$ is the extinction coefficient of the medium.

Assuming that all particles are significantly larger than the wavelength of light, equant, and randomly positioned and oriented, the mean particle volume is given as:
\begin{equation}
v = \frac{1-P}{n}.
\end{equation}

Hapke conceptualized the medium as a lattice of imaginary cubes, each with edge length $L$, which represents the mean center-to-center distance between particles:
\begin{equation}
\label{eq:L}
L = n^{-\frac{1}{3}}.
\end{equation}

Given the random positioning of particles, the probability of light transmission through several layers is the product of the probabilities for each layer. Thus, the fraction of light remaining after traversing a distance $t$ composed of $N = \frac{t}{L}$ layers is:
\begin{equation}
T(t) = \left(1 - E L\right)^N = e^{N \ln{\left(1 - E L\right)}} = e^{-KE t},
\end{equation}
where
\begin{equation}
\label{eq:HapkeK}
K = -\frac{\ln{\left(1 - E L\right)}}{E L}.
\end{equation}

The mean geometric cross-sectional area of a particle is given by:
\begin{equation}
\label{eq:sigma}
\sigma = \frac{E}{n}.
\end{equation}

Assuming the particles are equant, that the diameter of a sphere with the same volume as a particle equals the diameter of a sphere with the same geometric cross-sectional area, we have:
\begin{equation}
\label{eq:D}
D = \left(\frac{6}{\pi} v\right)^{\frac{1}{3}} = \left(\frac{4}{\pi} \sigma\right)^{\frac{1}{2}}.
\end{equation}

Using the equations above, $E L$ can be derived as follows:
\begin{equation}
\begin{aligned}
\label{eq:HapkeEL}
E L & = n \frac{E}{n} L \\
& = n^{\frac{2}{3}} \frac{E}{n} \\
& = \left(\frac{1-P}{v}\right)^{\frac{2}{3}} \frac{\pi D^2}{4} \\
& = \left(\frac{1-P}{\frac{\pi D^3}{6}}\right)^{\frac{2}{3}} \frac{\pi D^2}{4} \\
& = \frac{\left(1-P\right)^{\frac{2}{3}}}{D^2 \left(\frac{\pi}{6}\right)^{\frac{2}{3}}} \cdot \frac{\pi D^2}{4} \\
& = \frac{\pi}{4 \cdot \left(\frac{\pi}{6}\right)^{\frac{2}{3}}} \left(1-P\right)^{\frac{2}{3}} \\
& = \left(\frac{3\sqrt{\pi}}{4} \left(1-P\right)\right)^{\frac{2}{3}}.
\end{aligned}
\end{equation}

Substituting Eqn.~(\ref{eq:HapkeEL}) into Eqn.~(\ref{eq:HapkeK}), we obtain:
\begin{equation}
K = -\frac{\ln{\left(1 - \left(\frac{3\sqrt{\pi}}{4} \left(1-P\right)\right)^{\frac{2}{3}}\right)}}{\left(\frac{3\sqrt{\pi}}{4} \left(1-P\right)\right)^{\frac{2}{3}}}.
\end{equation}

\section{More Results}
\label{sec:results}

In this section, we provide additional results as mentioned in our main paper. The parameter settings for the scenes presented in the main paper are detailed in Table~\ref{tab:parameters}.

\begin{figure*}[t]
\centering
\includegraphics[width = 0.9\linewidth]{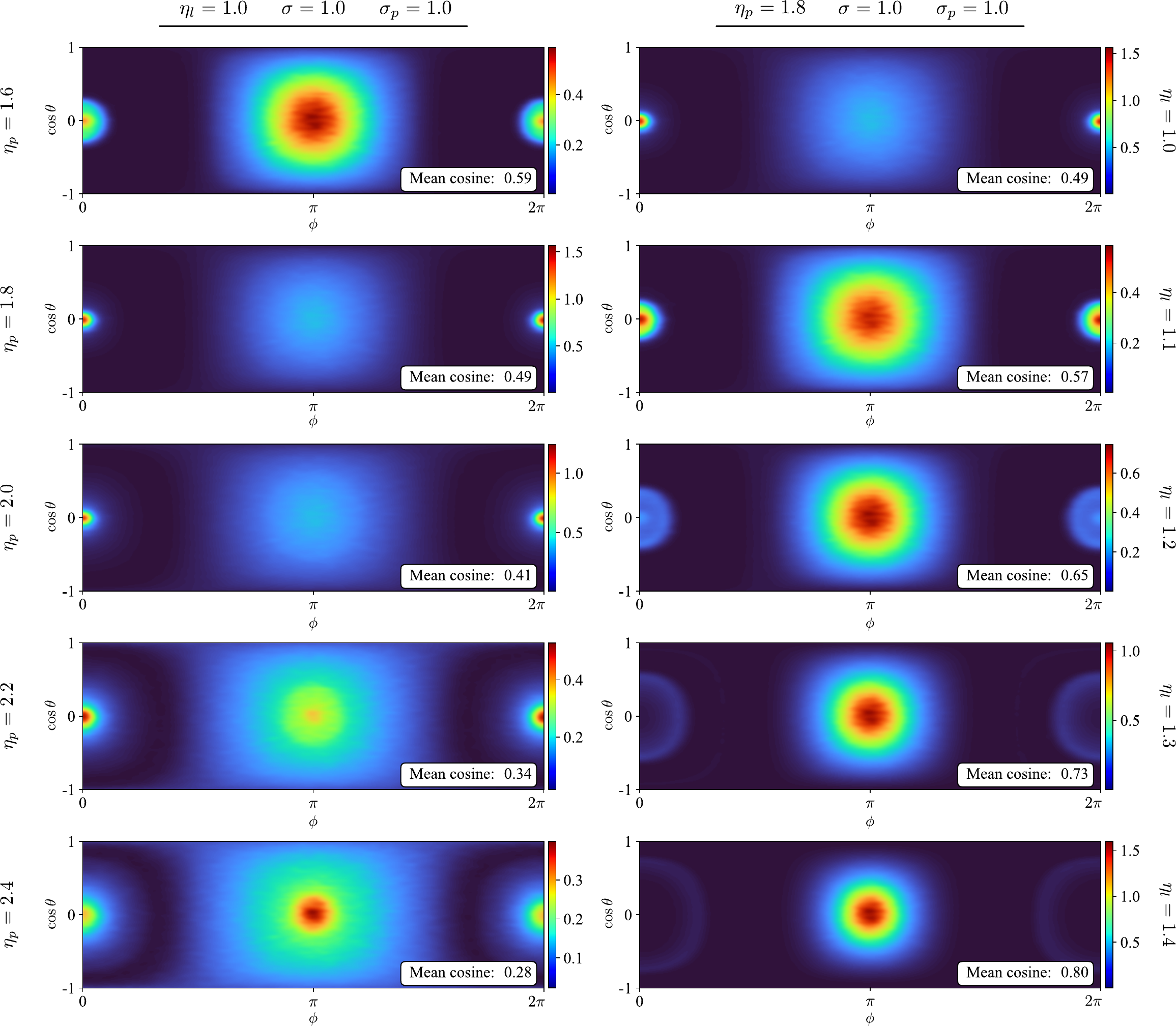}
\caption{Visualization of the
phase function of the anisotropic micro-volume model along with its corresponding mean cosine, over varying refractive index of particle $\eta_p$ and refractive index of liquid $\eta_l$.}
\label{fig:phaseCurve_ani_eta}
\end{figure*}

\begin{figure*}[t]
\centering
\includegraphics[width = 0.9\linewidth]{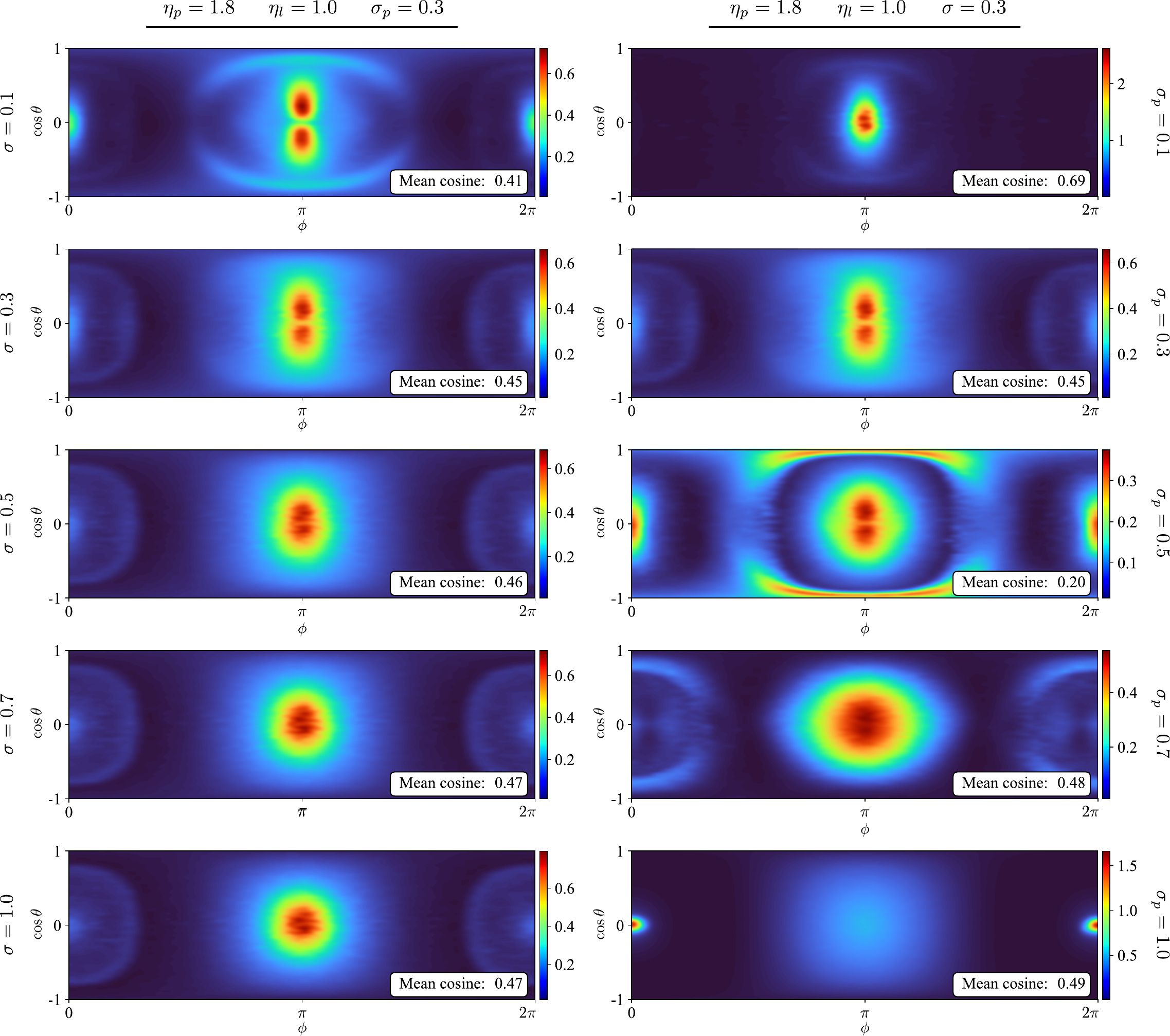}
\caption{Visualization of the
phase function of the anisotropic micro-volume model along with its corresponding mean cosine, over varying normal projected area $\sigma$ and particle projected area $\sigma_p$.}
\label{fig:phaseCurve_ani_sigma}
\end{figure*}

\paragraph{Phase function visualization}
We present a visualization of the phase function of the micro-ellipsoid model along with its corresponding mean cosine across different parameter settings in Fig.~\ref{fig:phaseCurve_ani_eta} and Fig.~\ref{fig:phaseCurve_ani_sigma}. 
% A phase function with a high $\eta_p$ exhibits a larger mean cosine, indicating stronger backward scattering. In contrast, a high $\eta_l$ results in a smaller mean cosine, corresponding to enhanced forward scattering, as we expected.

\begin{figure*}[t]
\centering
\includegraphics[width = 0.8\linewidth]{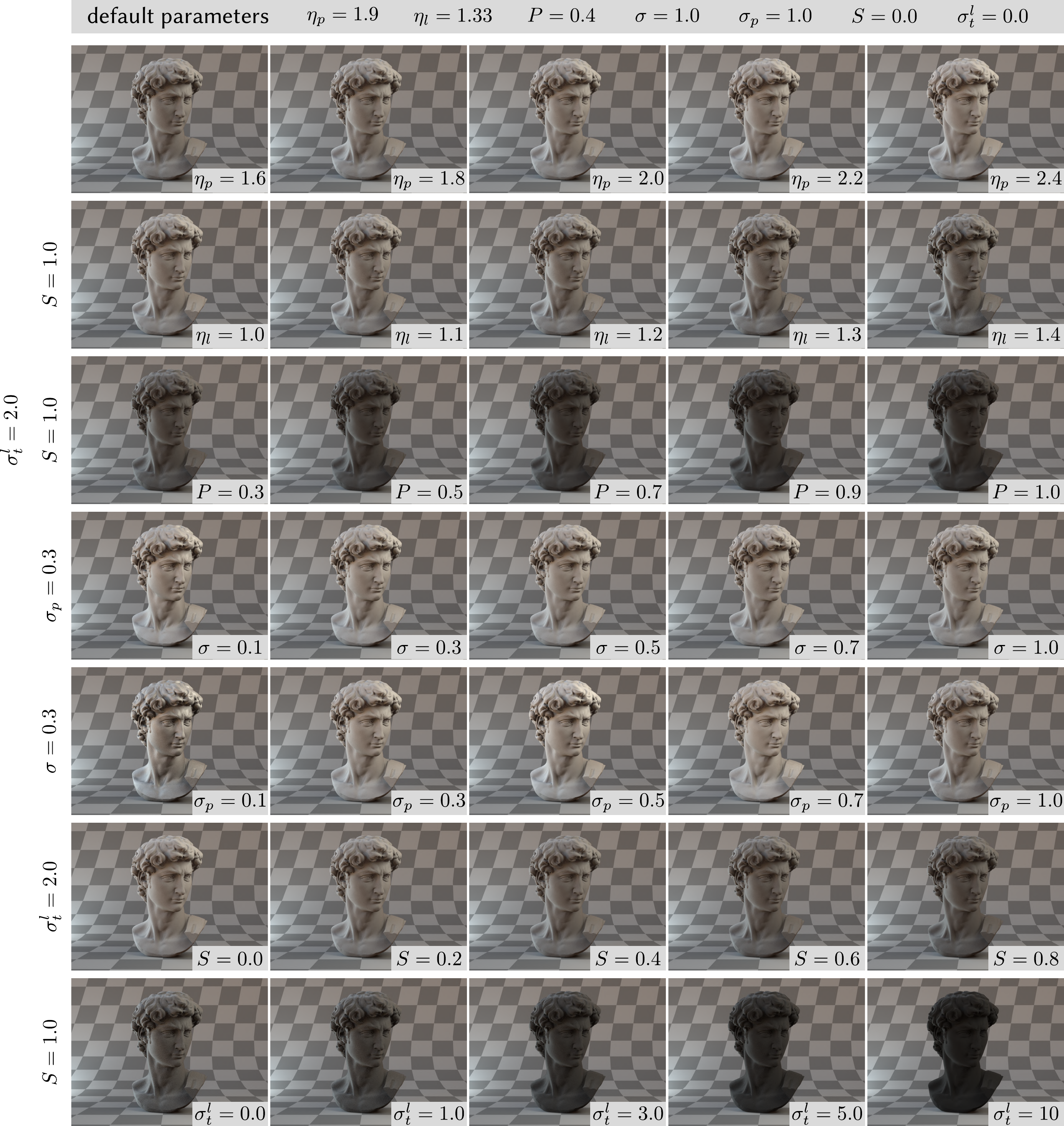}
\caption{Sculpture scene under different parameter settings. Parameters not explicitly mentioned are set to their default values, as indicated at the top of the figure.}
\label{fig:parametersRender}
\end{figure*}

\paragraph{Parameter analysis}
To demonstrate the impact of different parameters on appearance, we present renderings of the Sculpture scene under various parameter settings in Fig.~\ref{fig:parametersRender}. This scene is illuminated by two area light sources and an environment map. 

\begin{table*}[t]
\centering
   \resizebox{0.65\linewidth}{!}{
\begin{tabular}{llccccccccc}
\toprule
\textbf{Scene} & \textbf{Object}  & $P$ & $\alpha$ & $Z$ & $S$ & $\sigma_t^l$& $\eta_p$ & $\eta_l$ & $\sigma$ & $\sigma_p$ \\
\midrule
\multirow{3}{*}{Sand} & left & 0.425 & $(0.88, 0.83, 0.71)$ & $\infty$ & 0 & $(0,0,0)$ & 2.1 & 1 & 1 & 1\\
 & middle & 0.425 & $(0.88, 0.83, 0.71)$ & $\infty$ & 1 & $(0,0,0)$ & 2.1 & 1.33 & 1 & 1\\
 & right & 0.425 & $(0.88, 0.83, 0.71)$ & $\infty$ & 1 & $(1,2,2)$ & 2.1 & 1.33 & 1 & 1\\
\midrule
\multirow{1}{*}{Cloth} & cloth & 1 & $(0.65,0.09,0.18)$ & 4 & map & $(0,0,0)$ & 2.4 & 1.33 & 0.1 & 0.1\\
\midrule
\multirow{1}{*}{Paper} & paper & 0.5 & $(0.88, 0.90, 0.92)$ & 1 & map & $(0,0,0)$ & 1.7 & 1.33 & 1 & 1\\
\midrule
\multirow{1}{*}{Table} & table & 0.5 & map & $\infty$ & map & $(0,0,0)$ & 1.7 & 1.33 & 1 & 1\\
\midrule
\multirow{6}{*}{Teaser} & pot (left) & 0.5 & map & $\infty$ & map & $(0,0,0)$ & 1.8 & 1.33 & 1 & 1\\
& pot (right) & 0.5 & map & $\infty$ & map & $(1,1,1)$ & 1.8 & 1.33 & 1 & 1\\
& cloth (white) & 0.9 & $(0.7,0.7,0.7)$ & 1.5 & map & $(0,0,0)$ & 1.75 & 1.33 & 0.1 & 0.1\\
& cloth (red) & 0.9 & $(0.7,0.1,0.1)$ & $\infty$ & map & $(0,0,0)$ & 2.4 & 1.33 & 0.1 & 0.1\\
& sculpture (left) & 0.6 & $(0.85,0.85,0.85)$ & $\infty$ & map & $(0,0,0)$ & 1.8 & 1.33 & 1 & 1\\
& sculpture (right) & 0.6 & $(0.85,0.85,0.85)$ & $\infty$ & map & $(0,0,0)$ & 1.8 & 1.501 & 1 & 1\\
\bottomrule
\end{tabular}

}\caption{Parameters for different scenes and objects.}
\label{tab:parameters}
\end{table*}

% \clearpage 
\bibliographystyle{ACM-Reference-Format}
\bibliography{paper}